\documentclass[onecolumn,aps,prd,preprintnumbers,showpacs,superscriptaddress,nofootinbib,amsmath,amssymb,floats,floatfix,showkeys,notitlepage]{revtex4}

\usepackage{lipsum}
\usepackage{graphicx}
\usepackage{subfigure}
\usepackage{palatino}
\usepackage{changes}
\usepackage{hyperref}
\hypersetup{colorlinks=true,linkcolor=blue,urlcolor=blue,citecolor=blue}
\usepackage[toc,page]{appendix}
\usepackage[normalem]{ulem}
\usepackage{adjustbox}
\usepackage{latexsym}
\usepackage{amsmath}
\usepackage{amssymb}
\usepackage{amsfonts}
\usepackage{times}
\usepackage{dcolumn}
\usepackage{bm}
\usepackage{tikz}
\usepackage{bigints}
\usepackage{array,tabularx,multirow,booktabs}
\usepackage[tracking=true]{microtype}
\SetTracking{}{500}
\SetTracking{encoding={*}, shape=sc}{40}
\UseRawInputEncoding 
\allowdisplaybreaks

\begin{document} \sloppy
\title{Dehnen halo effect on a black hole in an ultra-faint dwarf galaxy}
\author{Reggie C. Pantig}
\email{reggie.pantig@dlsu.edu.ph}
\affiliation{Physics Department, De La Salle University, 2401 Taft Avenue, Manila, 1004 Philippines}
\author{Ali \"Ovg\"un}
\email{ali.ovgun@emu.edu.tr}
\homepage{https://aovgun.weebly.com/}
\affiliation{Physics Department, Eastern Mediterranean University, Famagusta, 99628 North Cyprus via Mersin 10, Turkey}

\begin{abstract}
There had been recent advancement toward the detection of ultra-faint dwarf galaxies, which may serve as a useful laboratory for dark matter exploration since some of them contains almost 99$\%$ of pure dark matter. The majority of these galaxies contain no black hole that inhabits them. Recently, there had been reports that some dwarf galaxies may have a black hole within. In this study, we construct a black hole solution combined with the Dehnen dark matter halo profile, which is commonly used for dwarf galaxies. We aim to find out whether there would be deviations relative to the standard black hole properties, which might allow determining whether the dark matter profile in an ultra-faint dwarf galaxy is cored or cuspy. To make the model more realistic, we applied the modified Newman-Janis prescription to obtain the rotating metric. We analyzed the black hole properties such as the event horizon, ergoregion, geodesics of time-like and null particles, and the black hole shadow. Using these observables, the results indicate the difficulty of distinguishing whether the dark matter is cored or cuspy. To find an observable that can potentially distinguish these two profiles, we also calculated the weak deflection angle to examine the effect of the Dehnen profile in finite distance and far approximation. Our results indicate that using the weak deflection angle is far better, in many orders of magnitude, in potentially differentiating these profiles. We conclude that although dwarf galaxies are dark matter-dominated places, the effect on the Dehnen profile is still dependent on the mass of the black hole, considering the method used herein.
\end{abstract}


\date{\today}

\maketitle

\section{Introduction} \label{int}
In the outskirts of the Milky Way galaxy, there are orbiting dwarf galaxies that home a thousand or a few billion stars. These dwarf galaxies are not created equal and are thought to originate during the collision of galaxies or through the earliest stages of galactic evolution of their parent galaxy. Hence, these are thought to contain crucial information about galaxy evolution since its physical form is similar to the earliest galaxies formed in the Universe. In galaxy evolution, dark matter plays a central role. Thus, dwarf galaxies may also contain a significant amount of dark matter in them and may serve as another area of study to uncover the mystery of dark matter, at least on a small scale \cite{Battaglia:2022dii}.

There are many types of dwarf galaxies, and this work is interested in a particular type called Ultra Faint Dwarf Galaxy or UFDs that can be as tiny as $30$ pc ($\sim 9.28\times10^{19}$ m) in half-light radius with a luminosity of around $L=10^{2-3}L_{\odot}$ \cite{Simon2017,Martin2016,Homma2016}. Detected UFDs are also confirmed to reside in the smallest dark matter halos yet found. Even so, they are the most dark matter-dominated systems in terms of the ratio between the halo mass and baryon mass, thus making it an excellent laboratory to explore the nature of dark matter \cite{Simon2019}. In this direction, there had been some debate as to whether the dark matter configuration in dwarf galaxies increases toward the galactic center \cite{Ishiyama2013} or appears to have a nearly constant dark matter density at central regions \cite{Hayashi2012}. If dark matter consists of warm self-interacting particles, it suggests that the central DM must be cored. However, since some UFDs contain $99\%$ dark matter \cite{Simon2019}, the baryonic effect may have some negligible influence on the central DM, suggesting that it must be cuspy due to the preservation of the primordial dark matter.

Building future facilities can be challenging in studying UFDs because of their unique properties, and future detectors may help determine whether the DM profile within is cored or cuspy. In addition, few studies have been made about the spectral properties of stars in UFDs that are used to measure their line-of-sight velocities, which could give rise to a DM profile model. Nevertheless, there are theoretical studies directed to alternatively find out whether the dark matter profile in UFDs is cored or cuspy. In Ref. \cite{Penarrubia2016}, tidal force disruption caused by different DM profile potentials for UFD was studied in a wide range of binary systems. Dynamical friction (DF) of a stellar distribution against a dark matter profile model has also been considered in this direction \cite{Hernandez2016} \cite{Inoue2017}. If the effect of DF is strong enough in the stellar distribution, the DM profile is a cuspy one, and a stellar cusp may form as well as a nucleus cluster. Otherwise, the DM profile is a cored one.

In this paper, we contribute to the mentioned literature by exploring the Dehnen \cite{Dehnen1993} dark matter profile influence on the known black hole properties as it admits both cored and cuspy configurations. Using the black hole, we will find out which configurations give a greater deviation. To do so, we will derive the new black hole metric with the Dehnen profile using the methods pioneered by Xu et al. \cite{Xu2018}, where the spin parameter $a$ of the black hole was also included using the Newman-Janis prescription \cite{Azreg-Ainou2014}. Such a method has been applied to analyze the effect of various dark matter profiles on Sgr. A* \cite{Hou_2018a}, and M87* \cite{Jusufi2019}. The effect of superfluid dark matter on Sgr. A* \cite{Jusufi2020} and Verlinde's Emergent Gravity effects on the black hole shadow have been analyzed \cite{Jusufi2021} using Xu et al. method. More complicated metrics incorporating dark matter profiles were also derived \cite{Xu2021a}, and dark matter spike (DM or GS spike) was also considered \cite{Xu2021b,Nampalliwar2021}. Another study of a black hole under the effect of dark matter distribution has also been studied in Ref. \cite{Konoplya2019,Pantig:2020uhp,Pantig:2021zqe,Konoplya:2022hbl}. Indeed, excitement in dark matter research has further elevated and become more significant due to important findings of the time evolution of the dark matter core as the unknown dark matter particles interact with baryonic matter \cite{Sharma2022}.

The Dehnen profile is commonly used in dwarf galaxies, and more often than not, these galaxies do not home a black hole at their centers. However, recent observations reveal that a massive black hole may also inhabit these dwarf galaxies. Specifically, it was reported in Ref. \cite{MkR2022} that an SMBH of mass $\sim2.00\text{x}10^5 M_{\odot}$ inhabits a dwarf galaxy Mrk 462. In addition, Ref. \cite{Schutte2022} studied a black hole-triggered star formation in dwarf galaxy Henize 2-10, where the mass is of the SMBH is $\sim1.00\text{x}10^6 M_{\odot}$. Finally, in Ref. \cite{Bustamante-Rosell:2021ldj}, a dynamical study of dark matter using photometric and spectroscopic data reveals a black hole in Leo I with a mass of $(3.3)\pm2\text{x}10^{6} M_\odot$, which is $13\%$ of Leo I galaxy's mass. Furthermore, the BH mass in Leo I is comparable to the mass of Sgr. A*. As we use the Dehnen profile, we need the unique values for the dark matter mass $k$ and core radius $r_\text{c}$ for this profile. While there was no available data for Mrk 462 and Henize 2-10, we find that for Leo I, $k = 2.00\text{x}10^{9}M_\odot$ and $r_\text{c} = 3.7$ kpc for the cuspy profile, and $k = 2.51\text{x}10^{10}M_\odot$ and $r_\text{c} = 10.7$ kpc for the cored profile (see Table $2$ of Ref. \cite{errani2018systematics}). In this paper, we will use Leo I and its black hole as a model to explore the effect of the Dehnen profile.

We will analyze in this paper the effect of the Dehnen profile on the event horizon, ergoregions, geodesics of time-like and null particles, and the black hole shadow. To further broaden the scope of this study, the calculation of the weak deflection angle will also be presented since it can also prove useful in probing dark matter behavior. We remark that the analysis using the weak deflection angle has never been done in the mentioned literature above due to the complicated metric resulting from Xu et al. method. However, this has been achieved in the recent paper in \cite{Pantig:2022toh} but in the non-rotating case. In this paper, the rotating case for the deflection angle will be considered. We will use the well-known geometrical technique that was developed by Gibbons and Werner \cite{Gibbons_2008} using the Gauss-Bonnet theorem (GBT) to the optical metric of asymptotically flat spacetime. It was further extended by Ishihara et al. \cite{Ishihara2016} to include non-asymptotically flat spacetime, giving a generalized expression concerning the finite positions of the source and the receiver. The axisymmetric version was developed in Ref. \cite{Ono:2019hkw}. There are various studies on asymptotic/non-asymptotic spacetimes, finite distance, and the use of GBT on black holes and wormholes, one can see \cite{Okyay:2021nnh,Jusufi:2017mav,Ovgun:2018tua,Ovgun:2018fnk,Ovgun:2019wej,Li:2020dln,Javed:2019rrg,Javed:2019ynm,Zhang:2021ygh,Belhaj:2020rdb,Kumar:2019pjp,Islam:2020xmy,Fu:2021akc}. Furthermore, using the photonsphere radius, \cite{Li:2020wvn} formulated a version of the GBT to include non-asymptotically flat spacetime. These generalized formulations are important in this study since the metrics consisting of dark matter profiles are proved to be complicated and non-asymptotically flat. Recent and related work for deflection angles caused by dark matter can be found in Refs. \cite{Metcalf2001,Ovgun2019,Ovgun2020,Atamurutov2022,Ovgun2019b}.

Before closing this section, we mentioned the DM spike, which has been considered in papers \cite{Nampalliwar2021,Xu2021b}. In general, other than the cuspy nature of the dark matter profile that is still being debated, DM spike occurs due to an assumption of adiabaticity of black hole growth that is ideally located at the center of the dark matter halo \cite{Ullio:2001fb}. Theoretically, the assumption of adiabaticity is still unverified, and weaker DM spikes (and cusps) occur when the SMBH is not at the center of the DM halo and when one considers the effects of dynamical heating. The non-existence of the stellar spike can also rule out DM spikes \cite{Lacroix:2018zmg}. In this paper, we do not consider the effect of DM spikes since we want to analyze the Dehnen profile in its simplest configuration for dwarf galaxies containing a black hole and located at the outer spiral arms of its parent galaxy. The Dehnen profile also offers analysis for both cored and cuspy configurations, where the cusp might be considered as a mini-DM spike.

The program of the paper is as follows: in Sect. \ref{sec2}, we derive the non-spinning black hole metric in the Dehnen profile. Next, we generalize this in Sect. \ref{sec3} using the modified Newman-Janis procedures to include the spin parameter $a$. In Sects. \ref{sec3}-\ref{sec8}, we study the effect of the Dehnen profile on some black hole properties such as the horizon, ergoregions, time-like and null geodesics, the black hole shadow, and the weak deflection angle. In Sect. \ref{sec9}, we state conclusive remarks and recommend some future research prospects. In this paper, we have used the natural units $G=c=1$, and signature ($-,+,+,+$).

\section{Black hole metric in Dehnen profile} \label{sec2}
In this article, we use the Dehnen profile presented in Ref. \cite{Dehnen1993,Stegmann2020}, where it was used to improve constraints from ultra-faint dwarf galaxies on primordial black holes as dark matter. The density profile reads
\begin{equation} \label{e1}
    \rho(r) = \frac{(3-\sigma)k}{4\pi r_{\text{c}}^3} \left(\frac{r}{r_{\text{c}}}\right)^{-\sigma} \left(1+\frac{r}{r_{\text{c}}}\right)^{\sigma-4}.
\end{equation}
Here, $k$ and $r_{\text{c}}$ are the dark matter halo's total mass and the scale radius of the dark matter halo, respectively. Such a profile is useful in describing an extremely low mass but compact ultra-faint dwarf galaxies surrounded by dark matter halo. We can find the mass profile at any radial distance $r$ of Eq. \eqref{e1} using the standard formula
\begin{equation} \label{e2}
    M_{\text{DM}}(r)=4 \pi \int_{0}^{r} \rho\left(r^{\prime}\right) r^{\prime 2} d r^{\prime},
\end{equation}
by which we obtain
\begin{equation} \label{e3}
    M_{\text{DM}}(r)=\frac{k\left[r^{3-\sigma}(r_{\text{c}}+r)^{\sigma}-\mathcal{Y}(r_{\text{c}}+r)^{3}\right]}{(r_{\text{c}}+r)^{3}},
\end{equation}
where
\begin{equation} \label{e4n}
	\mathcal{Y}=\lim_{r' \to 0^+} r'^{3-\sigma}(r_{\text{c}}+r')^{-3+\sigma}=0
\end{equation}
if $r_{\text{c}}>0$ and $\sigma<3$. Therefore, when Eq. \eqref{e4n} is used, the parameter $\sigma$ only makes sense if it is restricted in the interval $[0,3)$ \cite{Dehnen1993}. Since we are interested in the cored parameter $\sigma=0$ and the cuspy parameter $\sigma=1$, Eq. \eqref{e4n} should apply. Using $M_{\text{DM}}(r)$, we can find the tangential velocity associated with the profile for any test particle moving within the halo. Using the definition $v_{\text{t g}}^{2}(r)=M_{\text{DM}}(r) / r$, we find
\begin{equation} \label{e4}
    v_{\text{tg}}(r)=\sqrt{k}r^{1-\frac{\sigma}{2}}(r_{\text{c}}+r)^{\frac{1}{2}(\sigma-3)}.
\end{equation}
The line element describing a dark matter halo is given by
\begin{equation} \label{e5}
    ds^{2}_{\text{halo}} = -f(r) dt^{2} + g(r)^{-1} dr^{2} + r^2 d\theta ^{2} +r^2\sin^2\theta d\phi^{2}.
\end{equation}
Here the metric function $f(r)$ is related to $v_{\text{tg}}(r)$ through the expression
\begin{equation} \label{e6}
    v_{\text{tg}}(r)=r \frac{d \ln (\sqrt{f(r)})}{d r}.
\end{equation}
Solving for $f(r)$ yields
\begin{equation} \label{e7}
    f(r)=\exp\left[-\frac{2kr^{\sigma-2}(r_{\text{c}}+r)^{\sigma-2}}{r_{\text{c}}(\sigma-2)}\right].
\end{equation}
We can see that $\sigma \neq 2$ since $f(r)$ will be undefined. With the line element in Eq. \eqref{e5}, we see that a static and spherically symmetric DM halo restricts the parameter $\sigma$ to the interval $[0,2)$. Nonetheless, the expression is relevant for a non-isothermal cored dark matter halo when $\sigma = 0$, and a cuspy dark matter halo when $\sigma = 1$ \cite{Dehnen1993,Stegmann2020,Inoue2017}.

The aim is to combine the dark matter profile imprinted in $f(r)$ to the black hole metric function. To obtain the metric line element for such a case, we used the formalism developed by Xu et al. \cite{Xu2018}, which has been used recently by several authors \cite{Jusufi2019,Jusufi2020,Nampalliwar2021,Jusufi2021,Xu2021a,Xu2021b}. With the fusion of the black hole and dark matter halo, which are assumed to be static and spherically symmetric, one obtains the Einstein field equation of the form
\begin{equation} \label{e8}
    R^{\mu}_{\nu}=\frac{1}{2}\delta^{\mu}_{\nu}R=\kappa^2((T^{\mu}_{\nu})_{\text{DM}}+(T^{\mu}_{\nu})_{\text{Schw}}),
\end{equation}
which redefines the spacetime metric in Eq. \eqref{e5} as
\begin{equation} \label{e9}
    ds^{2} = -F(r) dt^{2} + G(r)^{-1} dr^{2} + r^2 d\theta ^{2} +r^2\sin^2\theta d\phi^{2},
\end{equation}
where
\begin{equation} \label{e10}
    F(r)=f(r) + F_1(r), \quad \quad G(r) = g(r)+F_2(r).
\end{equation}
As a consequence, Eq. \eqref{e8} gives us
\begin{align} \label{e11}
    (g(r)+F_{2}(r))\left(\frac{1}{r^{2}}+\frac{1}{r}\frac{g^{'}(r)+F^{'}_{2}(r)}{g(r)+F_{2}(r)}\right)&=g(r)\left(\frac{1}{r^{2}}+\frac{1}{r}\frac{g^{'}(r)}{g(r)}\right), \nonumber \\
    (g(r)+F_{2}(r))\left(\frac{1}{r^{2}}+\frac{1}{r}\frac{f^{'}(r)+F^{'}_{1}(r)}{f(r)+F_{1}(r)}\right)&=g(r)\left(\frac{1}{r^{2}}+\frac{1}{r}\frac{f^{'}(r)}{f(r)}\right).
\end{align}
After solving for $F_1(r)$ and $F_2(r)$ using the above equations, one finds that
\begin{align} \label{e12}
    F(r) &= \exp\left[\int \frac{g(r)}{g(r)-\frac{2m}{r}}\left(\frac{1}{r}+\frac{f^{'}(r)}{f(r)}\right)dr-\frac{1}{r} dr\right], \nonumber\\
    G(r) &=g(r)-\frac{2m}{r}.
\end{align}
If the dark matter halo is not considered, ie. $f(r)=g(r)=1$ since $k=0$, then the integral in $F(r)$ becomes a constant equal to $1-2m/r$, where $m$ is the black hole mass. Thus, we end up with the known Schwarzschild metric. With Eqs. \eqref{e11}-\eqref{e12}, we finally have the black hole metric with the dark matter halo. Lastly, under the assumption that $f(r)=g(r)$ and $F_1(r)=F_2(r)=-2m/r$, an immediate implication is that $F(r)=G(r)$ and the metric function $F(r)$ in Eq. \eqref{e12} can now be fully written as
\begin{equation} \label{e13} 
    F(r) = \exp\left[-\frac{2kr^{\sigma-2}(r_{\text{c}}+r)^{\sigma-2}}{r_{\text{c}}(\sigma-2)}\right]-\frac{2m}{r}.
\end{equation}
We observe that when the black hole is considered in the DM halo, the parameter $\sigma$ is again restricted to the interval $[0,2)$, and one has no choice but to only consider the cored or cuspy configurations. In the next sections, we want to change the notation for the metric line element in Eq. \eqref{e9} and write it as
\begin{equation} \label{e14}
    ds^{2} = -A(r) dt^{2} + B(r) dr^{2} + C(r) d\theta ^{2} +D(r) d\phi^{2},
\end{equation}
where $B(r)=A(r)^{-1}$, $C(r)=r^2$, and $D(r)=r^2\sin^2\theta$. With this notation, $A(r)=F(r)$, and $C(r)=D(r)$ when one analyzes the black hole properties along the equatorial plane where $\theta=\pi/2$.

\section{Rotating black hole metric in Dehnen profile} \label{sec3}
Realistic black holes are rotating with a certain spin parameter $a$. Thus, it is necessary to recast the metric to include $a$. With this aim, one popular but old method that we used is the Newman-Janis algorithm (NJA), which has been modified by the author in Ref. \cite{Azreg-Ainou2014} to only include a partial complexification procedure. With this modified version of the NJA, several authors have used this to obtain the rotating black hole metric that includes dark matter density profiles \cite{Xu2018,Jusufi2019,Jusufi2020,Pantig:2020uhp,Jusufi2021,Nampalliwar2021}. We ought to generalize the metric in Eq.\eqref{e14} by including the spin parameter. With this aim and following the NJA prescription, the first step is to convert the Boyer-Lindquist coordinates ($t,r,\theta,\phi$) into horizon penetrating coordinates ($u,r,\theta,\phi$), along with the use of the seed metric function $A(r)$ in Eq. \eqref{e13}:
\begin{equation} \label{eq15}
du=dt-dr^{*}=dt-\frac{dr}{A(r)},
\end{equation}
which enables us to rewrite the line element as
\begin{equation} \label{e16}
    d{s}^2 = -A(r)du^2 -2dudr +r^2d\theta^2 +r^2\sin^2 \theta d\phi^2.
\end{equation}
The line element can be written in terms of the null tetrads, where the contravariant metric tensor is expressed as
\begin{equation} \label{e17}
    {g}^{\mu\nu}=-{l}^\mu{n}^\nu-{l}^\nu{n}^\mu+{m}^\mu\bar{{m}}^\nu+{m}^\nu\bar{{m}}^\mu,
\end{equation}
and with the following definitions
\begin{align} \label{e18}
    &l=l^\mu\frac{\partial}{\partial x^\mu}=\delta^\mu_1\frac{\partial}{\partial x^\mu},\nonumber \\
    &n=n^\mu\frac{\partial}{\partial x^\mu}=\left(\delta^\mu_0-\frac{A(r)}{2}\delta^\mu_1\right)\frac{\partial}{\partial x^\mu},\nonumber \\
    &m=m^\mu\frac{\partial}{\partial x^\mu}=\frac{1}{\sqrt{2C(r)}}\left(\delta^\mu_2+\frac{i}{\sin\theta}\delta^\mu_3\right)\frac{\partial}{\partial x^\mu},\nonumber \\
    &\bar{m}=\bar{m}^\mu\frac{\partial}{\partial x^\mu}=\frac{1}{\sqrt{2C(r)}}\left(\delta^\mu_2-\frac{i}{\sin\theta}\delta^\mu_3\right)\frac{\partial}{\partial x^\mu},
\end{align}
where the null tetrad vectors satisfy normalization, orthogonality, and isotropy, ie.
\begin{align} \label{e19}
    l^\mu l_\mu&=n^\mu n_\mu=m^\mu m_\mu=\bar{m}^\mu \bar{m}_\mu=0 \nonumber\\
    l^\mu m_\mu&=l^\mu \bar{m}_\mu=n^\mu m_\mu=n^\mu \bar{m}_\mu=0
    \nonumber\\
    -l^\mu n_\mu&=m^\mu \bar{m}_\mu = 1.
\end{align}
The next step is to perform a basic coordinate complex transformation by writing
\begin{equation} \label{e20}
    {x'}^{\mu} = x^{\mu} + ia (\delta_r^{\mu} - \delta_u^{\mu})
    \cos\theta \rightarrow \\ \left\{\begin{array}{ll}
    u' = u - ia\cos\theta, \\
    r' = r + ia\cos\theta, \\
    \theta' = \theta, \\
    \phi' = \phi \end{array}\right.
\end{equation}
where $a$ is the spin parameter
The known metric coefficients are now then assumed to transform from $({A(r),B(r),C(r))}$ to an unknown coefficients $(\mathcal{A}(r,\theta,a),\mathcal{B}(r,\theta,a),\mathcal{C}(r,\theta,a))$. With the transformation of the null tetrad vector components via
\begin{equation} \label{e21}
    \left({l^\prime}^\mu,{n^\prime}^\mu,{m^\prime}^\mu,{\bar{m^\prime}}^\mu\right)=\frac{{\partial{x^\prime}^\mu}}{{\partial x^\nu}}{e_a}^\nu=\left(\begin{array}{cccc}
1 & 0 & ia\sin\theta & 0\\
0 & 1 & -ia\sin\theta & 0\\
0 & 0 & 1 & 0\\
0 & 0 & 0 & 1
\end{array}\right){e_a}^\nu,
\end{equation}
we can see that \cite{Azreg-Ainou2014}
\begin{align} \label{e22}
    &l^{\prime\mu}=\delta^\mu_1, \nonumber \\ 
    &n^{\prime\mu}=\left(\sqrt{\frac{\mathcal{B}}{\mathcal{A}}}\delta^\mu_0-\frac{\mathcal{B}}{2}\delta^\mu_1\right),\nonumber \\
    &m^{\prime\mu}=\frac{1}{\sqrt{2\mathcal{C}}}\left[\left(\delta^\mu_0-\delta^\mu_1\right)ia\sin\theta+\delta^\mu_2+\frac{i}{\sin\theta}\delta^\mu_3\right].
\end{align}
Using these transformed null tetrads, the new inverse metric coefficients are
\begin{equation} \label{e23}
{g^\prime}^{\mu\nu}=\left(\begin{array}{cccc}
\frac{a^{2}\sin^{2}\theta}{\mathcal{C}} & -\sqrt{\frac{\mathcal{B}}{\mathcal{A}}}-\frac{a^{2}\sin^{2}\theta}{\mathcal{C}} & 0 & \frac{a}{\mathcal{C}}\\
-\sqrt{\frac{\mathcal{B}}{\mathcal{A}}}-\frac{a^{2}\sin^{2}\theta}{\mathcal{C}} & \mathcal{B}+\frac{a^{2}\sin^{2}\theta}{\mathcal{C}} & 0 & -\frac{a}{\mathcal{C}}\\
0 & 0 & \frac{1}{\mathcal{C}} & 0\\
\frac{a}{\mathcal{C}} & -\frac{a}{\mathcal{C}} & 0 & \frac{1}{\mathcal{C}\sin^{2}\theta}
\end{array}\right).
\end{equation}
Noting that ${g\prime}_{\mu\nu}=({g^\prime}^{\mu\nu})^{-1}$, and as pointed out in Ref. \cite{Azreg-Ainou2014} that if one considers $\mathcal{K}(r)=C(r)$, and $B(r)=1/A(r)$ in a static and spherically symmetric seed metric, then
\begin{align} \label{e24}
    \mathcal{A}(r,\theta)&=\frac{(A(r)C(r)+a^2\cos^2\theta)\mathcal{C}(r,\theta)}{(C(r)+a^2\cos^2\theta)^2}, \nonumber\\
    \mathcal{B}(r,\theta)&=\frac{(A(r)C(r)+a^2\cos^2\theta)}{\mathcal{C}(r,\theta)}, \nonumber\\
    \mathcal{C}(r,\theta)&= r^2+a^2\cos^2.
\end{align}
Now from the Eddington-Finkelstein coordinates, one can use the coordinate transformation
\begin{equation} \label{e25}
du=dt-\frac{{r}^2+a^2}{\Delta(r)}dr, \quad d\phi=d\phi^\prime-\frac{a}{\Delta(r)}dr
\end{equation}
to go back to the Boyer-Lindquist coordinates after dropping the prime in the second expression. Here,
\begin{equation} \label{e26}
    \Delta(r)=r^2A(r)+a^2.
\end{equation}
It is useful to express Eq. \eqref{e26} into a familiar Kerr-like form by introducing
\begin{equation} \label{e27}
    \mathcal{M}(r)=\frac{r(1-A(r))}{2}
\end{equation}
so that
\begin{equation} \label{e28}
    \Delta(r)=r^2-2\mathcal{M}(r) r+a^2.
\end{equation}
We then obtained the rotating black hole metric in a Dehnen dark matter halo:
\begin{equation} \label{e29}
    ds^2=-\left(1-\frac{2\mathcal{M}(r) r}{\Sigma}\right)dt^2 - \frac{4\mathcal{M}(r)ar\sin^2\theta}{\Sigma}dtd\phi + \frac{\Sigma}{\Delta}dr^2 + \Sigma d\theta^2 + \frac{\left[(r^2+a^2)^2-a^2\Delta\sin^2\theta\right]}{\Sigma}\sin^2\theta d\phi^2.
\end{equation}
We can see that Eq. \eqref{e14} is modified by adding the cross-term $dtd\phi$ whose coefficient is now defined as \cite{Ono:2019hkw}
\begin{equation} \label{e30}
    2H(r)=\frac{4\mathcal{M}(r)ar\sin^2\theta}{\Sigma}.
\end{equation}
Furthermore, $B(r)$ is no longer equal to the inverse of $A(r)$ in the spinning black hole metric.

\section{Horizon and Ergoregion} \label{sec4}
\begin{figure}
    \centering
    \includegraphics[width=0.60\textwidth]{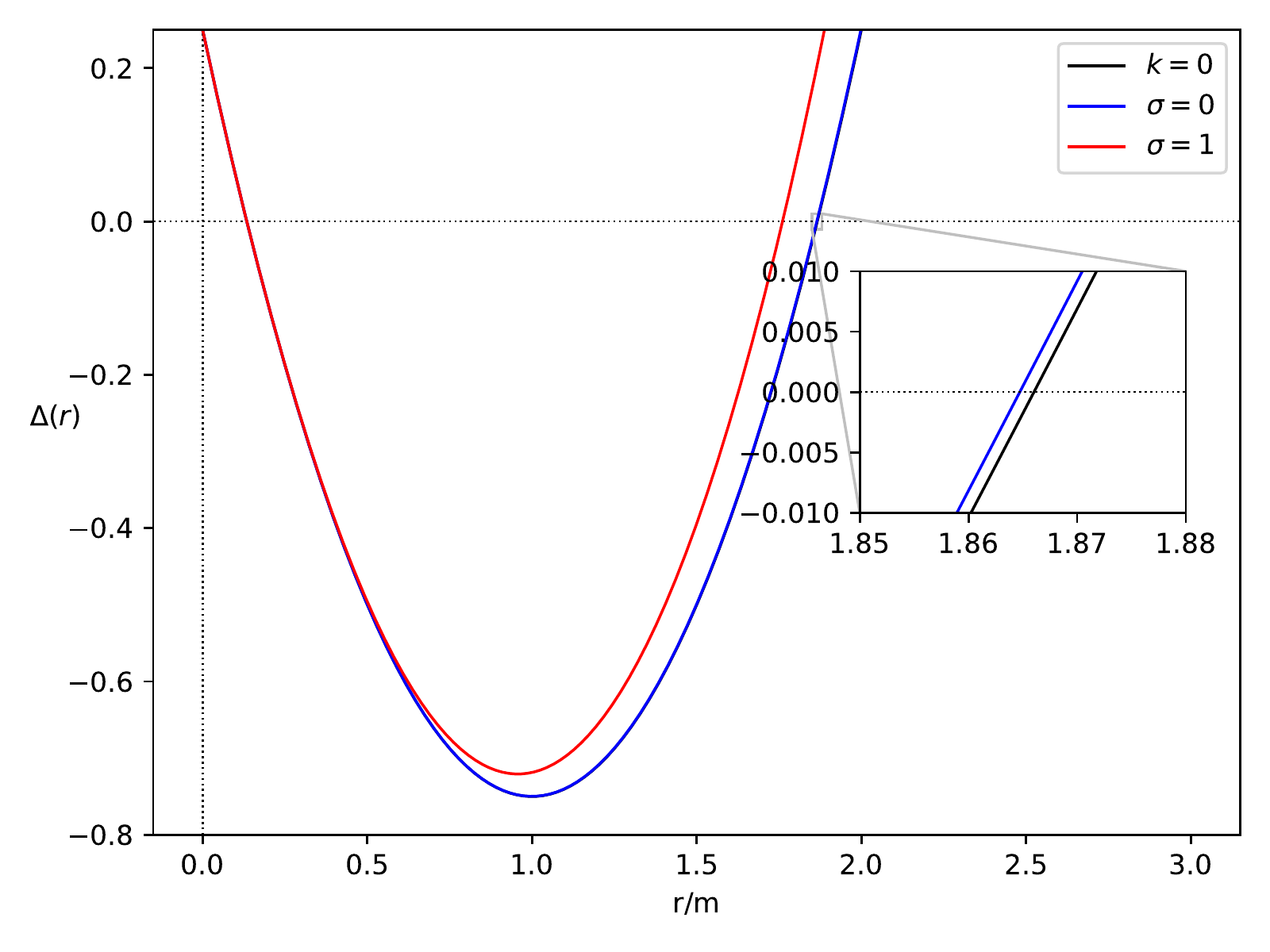}
    \caption{Plot of $\Delta(r)=0$ showing the location of the horizons formed when $a=0.50m$. There is still a difference seen between $k=0$, and $\sigma=0$.}
    \label{fig1}
\end{figure}

Let us now examine how the horizon and the ergoregion of the rotating black hole change due to the presence of the dark matter described by the Dehnen profile. Our program is to examine the effects of the Dehnen profile theoretically by means of numerical plots, then apply the situation to one of the observed dwarf galaxies that homes a black hole. Here, we will use the black hole recently discovered in the dwarf galaxy Leo I \cite{Bustamante-Rosell:2021ldj}, where $m=3.3\text{x}10^{6} M_\odot$. At this time of writing, however, there is no data about the value of the BH's spin parameter. Speculating a bit since this black hole is comparable to the mass of Sgr. A*, we will align its spin parameter to that of Sgr. A*, which only has a moderate positive/prograde spin of $a = 0.50m$ \cite{EventHorizonTelescope:2022xnr} at lower inclination angle. Furthermore, we will use the following Dehnen parameters for this dwarf galaxy : for cored type ($\sigma = 0$), $k = 2.51\text{x}10^{10}M_\odot$ and $r_\text{c} = 10.7$ kpc. For cuspy type ($\sigma = 1$), $k = 2.00\text{x}10^{9}M_\odot$ and $r_\text{c} = 3.7$ kpc \cite{errani2018systematics} . Inspecting Eq. \eqref{e29}, the locations of the horizons can be found by
\begin{equation} \label{e31}
    \Delta(r)=r^2-2\mathcal{M}(r) r+a^2=0.
\end{equation}
However, due to Eq. \eqref{e28}, we cannot simply solve Eq. \eqref{e31} analytically. Fig. \ref{fig1} shows the numerical plot of the horizons formed when $a=0.50m$. Here, we assumed theoretical values for $k$ and $r_c$. In particular, $k = 100m$, and $r_c=80m$ such that the halo is concentrated near the black hole. Although exaggerated, we can see the Dehnen effects in action \cite{Konoplya2019}. We can see then that the Dehnen profile decreases the radius of the event horizon with the cuspy profile $\sigma=1$ showing more deviation than the cored profile $\sigma=0$ relative to the Kerr case $k=0$. Remarkably, the Cauchy horizon has a vanishingly small deviation due to the Dehnen profile. Now, using the black hole in dwarf galaxy Leo I, the result for the value of the inner and outer horizons are shown in Table \ref{Tab1}.
\begin{table} [!ht]
    \centering
    \begin{tabular}{ |c|c|c| }
    \hline
    Profile &  Cauchy horizon & Event horizon \\
    \hline
    no DM ($k=0$) & 0.13397459621556135324  & 1.86602540378443864680   \\
    cored ($\sigma=0$) & 0.13397459621556135324 & 1.86602540378443864680  \\
    cuspy ($\sigma=1$) & 0.13397459621556135324  & 1.86602540378443863850   \\
    \hline
    \end{tabular}
    \caption{Values of the inner and outer horizons of a black hole in Leo I immersed in dark matter halo described by the Dehnen profile.}
    \label{Tab1}
\end{table}
While the deviation in the Cauchy horizon seems to be unaffected, the deviation seen in the event horizon radius for these cases is almost negligible. Interestingly, we can see the deviation in the cuspy profile around $10^{-17}$ order of magnitude. The deviation for the cored profile must be way smaller than this.

\begin{figure}
    \centering
    \includegraphics[width=0.60\textwidth]{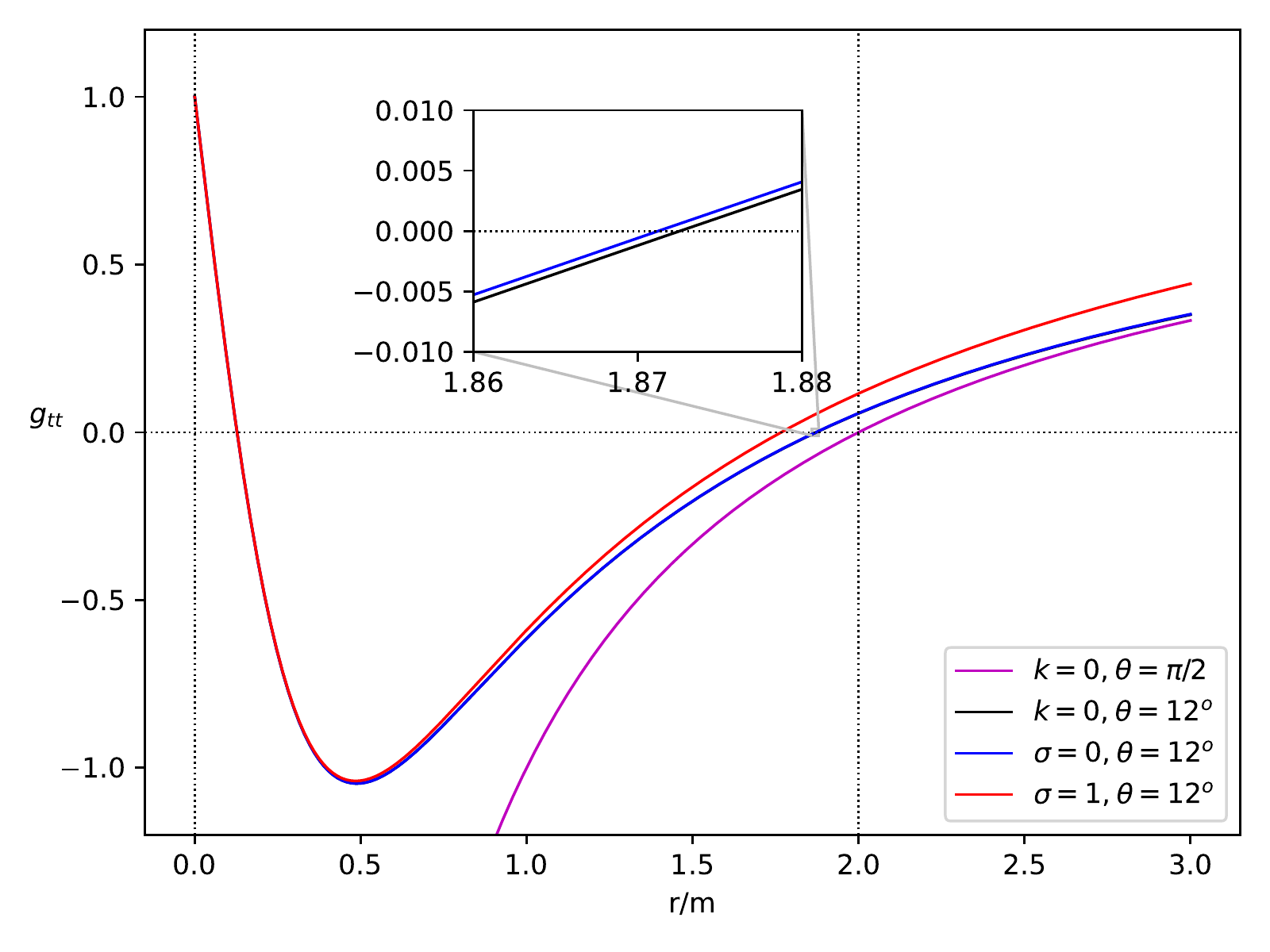}
    \caption{Plot of $g_\text{tt}=0$ showing the location of the ergoregions when $a=0.50m$. There is still a difference seen between $k=0$, and $\sigma=0$.}
    \label{fig2}
\end{figure}

The ergoregion, which is the region of static limit, can be found by solving $r$ in the equation
\begin{equation} \label{e32}
    \left(1-\frac{2\mathcal{M}(r) r}{\Sigma}\right)=0.
\end{equation}
In the equatorial plane where $\theta=\pi/2$, Eq. \eqref{e32} is independent of $a$, hence, a single ergoregion forms at $r=2m$. Analyzing the ergoregion when $\theta = 12.31^o$ \cite{NASA} for Leo I, two ergoregions form, and the Dehnen profile's effect is to decrease its outer radius. Similar to the Cauchy horizon, the inner ergo region has negligible deviation. We remark that the cuspy profile shows more deviation than the cored profile (see Fig. \ref{fig2}), where a very tiny deviation only occurs relative to the Kerr case.
\begin{table} [!ht]
    \centering
    \begin{tabular}{ |c|c|c| }
    \hline
    Profile &  Inner ergosphere & Outer ergosphere \\
    \hline
    no DM ($k=0$) & 0.12744083975273268748  & 1.87255916024726731250   \\
    cored ($\sigma=0$) & 0.12744083975273268748 & 1.87255916024726731250  \\
    cuspy ($\sigma=1$) & 0.12744083975273268748  & 1.87255916024726730420   \\
    \hline
    \end{tabular}
    \caption{Values of the inner and outer ergoregions of a black hole in Leo I immersed in dark matter halo described by the Dehnen profile.}
    \label{Tab2}
\end{table}
In Table \ref{Tab2}, the parameters used for Leo I reveal deviation only for the cuspy profile at around $10^{-17}$ order of magnitude.

\section{Time-like geodesics} \label{sec5}
In this section, we first analyze time-like geodesics. In particular, we will examine the effective potential and the innermost stable circular orbit (ISCO). To do so, we will apply the Hamilton-Jacobi approach expressed as
\begin{equation} \label{e33}
\frac{\partial S}{\partial \lambda}=-H,
\end{equation}
where $S$ is the action that is defined in terms of an affine parameter $\lambda$ and coordinates $x^{\mu }$. In General Relativity, the Hamiltonian takes the form
\begin{equation} \label{e34}
H=\frac{1}{2}g^{\mu \nu }\frac{
\partial S}{\partial x^{\mu }}\frac{\partial S}{\partial x^{\nu }},
\end{equation}
and using Eq. \eqref{e33}, it follows that
\begin{equation} \label{e35}
\frac{\partial S}{\partial \lambda }=-\frac{1}{2}g^{\mu \nu }\frac{
\partial S}{\partial x^{\mu }}\frac{\partial S}{\partial x^{\nu }}.
\end{equation}
Using the anzats
\begin{equation} \label{e36}
S=\frac{1}{2}\mu ^{2}\lambda -Et+L\phi +S_{r}(r)+S_{\theta }(\theta)
\end{equation}
due to the variable independence in $t$, $\phi$, and $\lambda$, the following equations of motion can be derived:
\begin{align} \label{e37}
&\Sigma\frac{dt}{d\lambda}=\frac{r^2+a^2}{\Delta(r)}P(r)-a(aE\sin^2\theta-L), \nonumber \\
&\Sigma\frac{dr}{d\lambda}=\sqrt{R(r)}, \nonumber \\
&\Sigma\frac{d\theta}{d\lambda}=\sqrt{\Theta(\theta)}, \nonumber \\
&\Sigma\frac{d\phi}{d\lambda}=\frac{a}{\Delta(r)}P(r)-\left(aE-\frac{L}{\sin^2\theta}\right),
\end{align}
with $P(r)$, $R(r)$ and $\Theta(\theta)$ given by
\begin{align} \label{e38}
&P(r)=E(r^2+a^2)-aL,\nonumber \\
&R(r)=P(r)^2-\Delta(r)[Q+(aE-L)^2+\mu ^2r^2], \nonumber \\
&\Theta(\theta)=Q-\left[a^{2}\left(\mu^{2}-E^{2}\right)+\frac{L^{2}}{\sin^{2}\theta}\right] \cos^2\theta,
\end{align}
with $Q$ being the Carter constant: $Q\equiv K-(L-aE)^{2}$ and $K$ is another constant of motion. Here, $\mu $ is proportional to the particle's rest mass.

\begin{figure}
    \centering
    \includegraphics[width=0.60\textwidth]{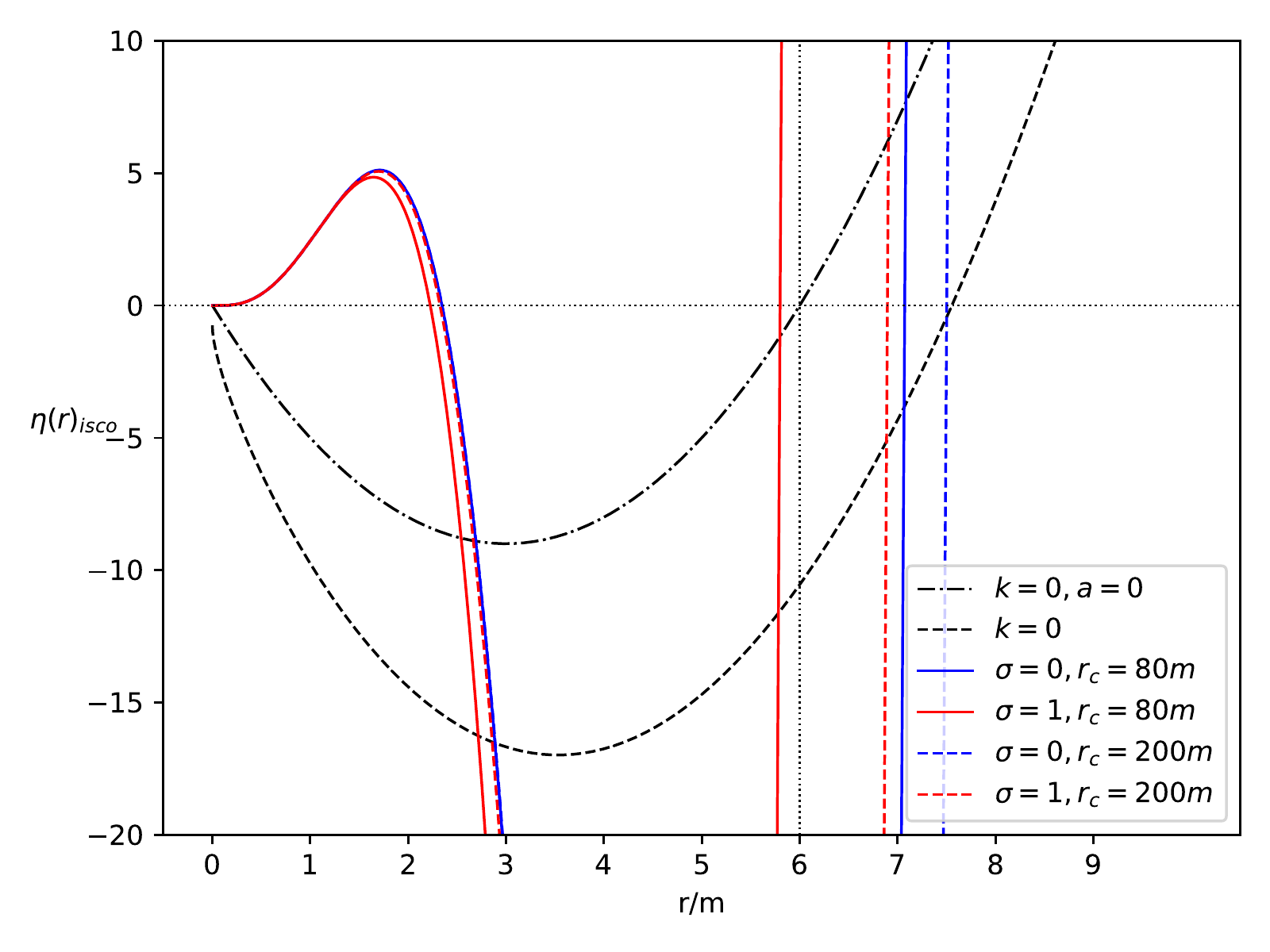}
    \caption{Plot of $\eta(r)_\text{isco}=0$ for the case of a massive particle that counter-rotates with the black hole (retrograde orbit). Two ISCOs are formed due to the effect of the Dehnen profile.}
    \label{fig3}
\end{figure}

In studying circular orbits, the condition
\begin{equation} \label{e39}
R(r)=\frac{dR(r)}{dr}\mid_{r=r_{\text{o}}}=0
\end{equation}
must be satisfied. Furthermore, the second derivative of $R(r)$ gives information about the stability of the circular orbit. With the aim of finding the location of the ISCO, we first find the energy $E_{\text{cir}}$ due to circular motion, where we use Eq. \eqref{e39}. Assigning $X^2=L-aE$ \cite{Slany2013}, we find
\begin{equation} \label{e40}
    X^2=\frac{r^3 \left(\Delta '(r)-2 E_{\text{cir}} ^2 r\right)}{-2 a^2-r \Delta '(r)+2 \Delta (r)}.
\end{equation}
Now, the energy $E_{\text{cir}}=E_{\text{isco}}$ if we take differentiate again Eq. \eqref{e40} with respect to $r$. After some considerable algebra,
\begin{equation} \label{e41}
    E_\text{isco}^2=\frac{1}{B r}\biggl\{a^2 \left[-\left(r \Delta ''(r)+3 \Delta '(r)\right)\right]+r \Delta (r) \Delta ''(r)-2 r \Delta '(r)^2+3 \Delta (r) \Delta '(r)\biggr\}
\end{equation}
where $B=-8 a^2+r \left(r \Delta ''(r)-5 \Delta '(r)\right)+8 \Delta (r)$, and finally the ISCO radius can be found via \cite{Pantig:2020uhp}
\begin{align} \label{e42}
    \eta(r)_{\text{isco}}&=\pm2 \Delta (r) \left(a^2-\Delta (r)\right)^2\pm\frac{9}{4} r \Delta (r) \left(a^2-\Delta (r)\right) \Delta '(r)
    \pm\frac{1}{16} r^3 \Delta '(r) \left(\Delta (r) \Delta ''(r)-2 \Delta '(r)^2\right)\nonumber \\
    &\pm\frac{1}{16} r^2 \bigl[4 \Delta (r) \left(a^2-\Delta (r)\right) \Delta ''(r)+\left(15 \Delta (r)-4 a^2\right) \Delta '(r)^2\bigr] \nonumber \\
    &+a \Delta (r) \sqrt{4 a^2+2 r \Delta '(r)-4 \Delta (r)} \bigl[-a^2+\frac{1}{8} r \bigl(r \Delta ''(r)-5 \Delta '(r)\bigr)+\Delta (r)\bigr]=0,
\end{align}
which reduces to the Kerr case
\begin{equation} \label{e43}
    3 a^2\mp8 a \sqrt{m} \sqrt{r}+r (6 m-r)=0
\end{equation}
if $\Delta(r)=r^2-2mr+a^2$. The upper sign in Eq. \eqref{e42} gives the particle prograde orbit, while the lower sign gives the retrograde orbit. Eq. \eqref{e42} seems a formidable equation, especially how $\Delta(r)$ is defined in Eq. \eqref{e28}. Hence, we use numerical plotting to Eq. \eqref{e42}, and Fig. \ref{fig3} reveals the location of the ISCO radii for the Schwarzschild, Kerr, and Kerr with the Dehnen profile. Note that the circular orbit of a massive particle should be at $r=6m$ for the Schwarzschild case. For the Kerr case, we consider first particles that counter-rotate with the black hole. These are particles with $L<0$. See Fig. \ref{fig3}. We see the expected ISCO radius for the Kerr case when $a=0.50m$, which is $r ~ 7.55m $. If the case is extremal ($a = m$) this is $r = 9m$. We could see only one ISCO for the Schw and Kerr cases. We can see that the Dehnen profile causes another ISCO to be formed near the horizon, which is related to another value of the angular momentum $L$ of the massive particle. The existence of multiple ISCO radii is not new. For example, when the spin of a particle is considered orbiting around a charged, non-rotating black hole, there exists two innermost stable circular orbits \cite{Zhang2019}. For the Kerr case, multiple ISCOs for a spinning particle are also found \cite{Jefremov2015}. The plot also shows that the deviation of the ISCO near the horizon is tiny compared to the second ISCO. For the second ISCO, denser dark matter distribution causes more deviation, as shown by the solid lines. In general, the difference between the cored and cuspy are evident, at least in the second ISCO.

\begin{figure}
    \centering
    \includegraphics[width=0.60\textwidth]{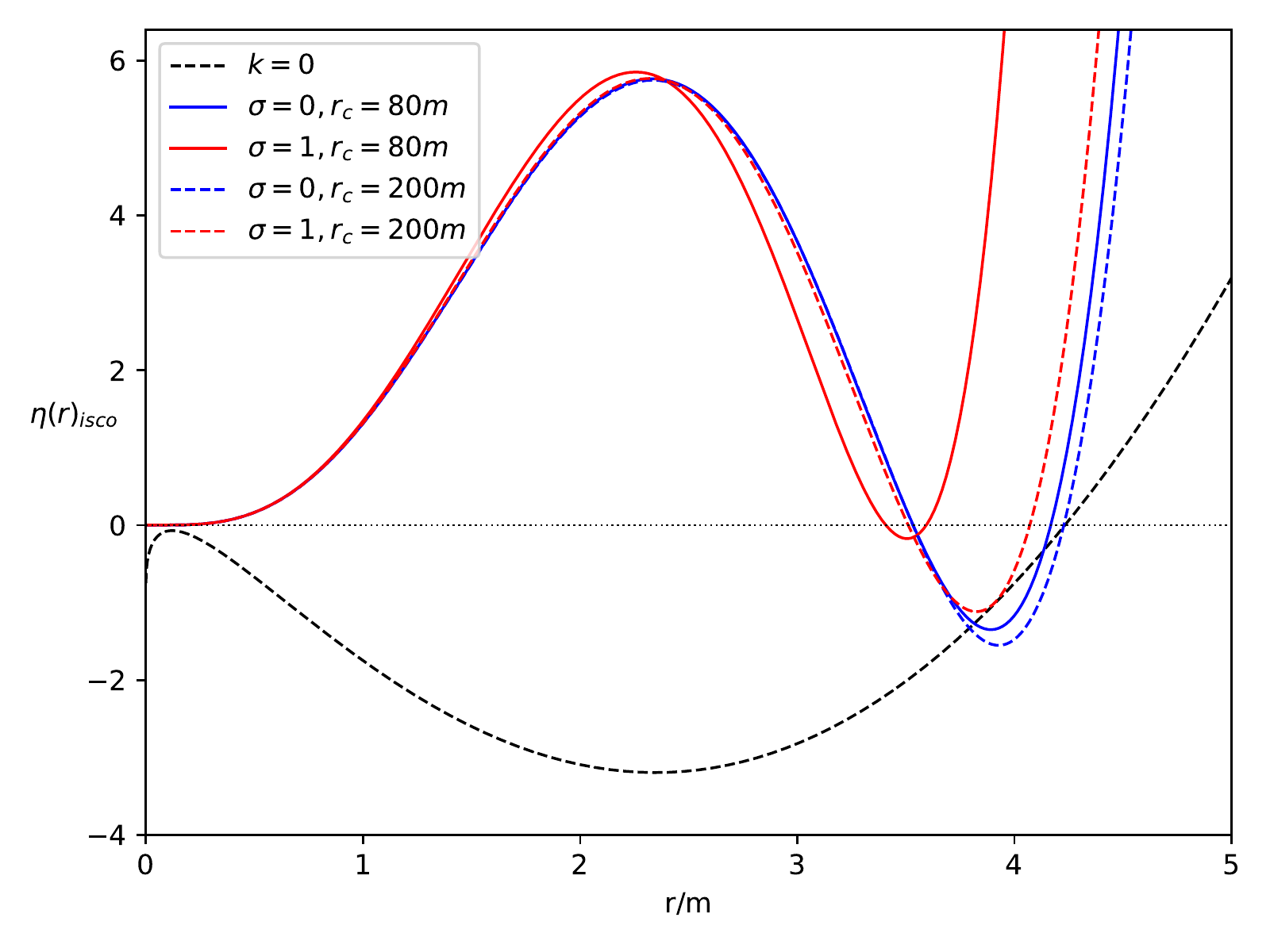}
    \caption{Plot of $\eta(r)_\text{isco}=0$ for the case of a massive particle that co-rotates with the black hole (prograde orbit). Two ISCOs are formed due to the effect of the DM profile.}
    \label{fig4}
\end{figure}

In Fig. \ref{fig4}, we also plotted the case where the massive particle co-rotates with the black hole. Note that in the extreme Kerr case, this ISCO has a value of $r=m$. For our case where $a = 0.50m$, $r=4.23m$. Due to the effect of the Dehnen profile, two ISCO radii are also formed, one found at a larger radius. Similar to the retrograde case, the dark matter effect tends to decrease the ISCO radius. Considering the deviation in the first ISCO, the cuspy profile (solid red line) seems to show a clear deviation from the other cases, unlike in the retrograde case. The cored and cuspy configuration can be discerned clearly in the second ISCO if the dark matter density is high (solid lines). Applying the Dehnen profile to Leo I, we tabulate the locations of the ISCO. Indeed, dark matter could be abundant in dwarf galaxies, but the black hole in Leo I is massive enough to counteract the Dehnen profile's effect, resulting in a similar situation where the dark matter density is so diluted. For the prograde case, the deviation can only be seen within $10^{-16}$ order of magnitude, while in the retrograde case, around $10^{-15}$ (see Table \ref{Tab3}). Remarkably, this implies that massive particles are sensitive to dark matter effects, especially at a far distance from the black hole.
\begin{table} [!ht]
    \centering
    \begin{tabular}{ |c|c|c| }
    \hline
    Profile &  Prograde & Retrograde \\
    \hline
    no DM ($k=0$) & 4.23300252953082567730  & 7.55458471451235844170   \\
    cored ($\sigma=0$) & 4.23300252953082567730 & 7.55458471451235844170  \\
    cuspy ($\sigma=1$) & 4.23300252953082559070  & 7.55458471451235796180   \\
    \hline
    \end{tabular}
    \caption{ISCO radii due to a black hole in Leo I immersed in dark matter halo described by the Dehnen profile.}
    \label{Tab3}
\end{table}

\begin{figure}
    \centering
    \includegraphics[width=0.60\textwidth]{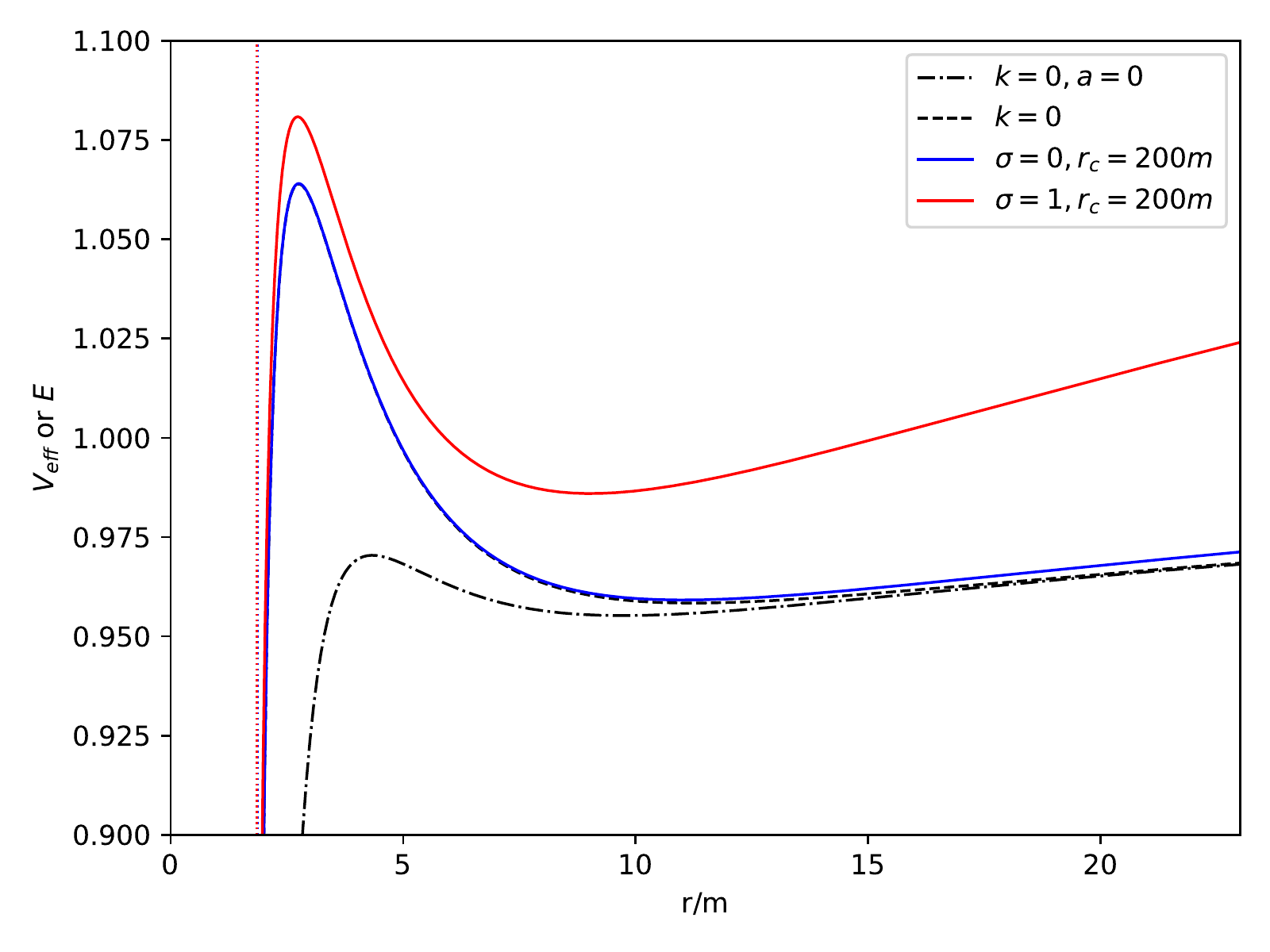}
    \caption{The effective potential when $a=0.50m$ for $+L$. The value of the angular momentum per unit mass is $L=3.75m$.}
    \label{fig5}
\end{figure}

We now focus on the massive particle's effective potential, where we can obtain qualitative impressions about the bound, stable, and unstable circular orbits. The effective potential in terms of angular momentum per unit mass $L$, given in terms of the inverse metric of Eq. \eqref{e29} is \cite{Olvera2019}
\begin{equation} \label{e44}
    V_{\pm}=\frac{g^{t\phi}}{g^{tt}}L\pm \Bigg\{\left[\left(\frac{g^{t\phi}}{g^{tt}}\right)^2-\frac{g^{\phi \phi}}{g^{tt}}\right]L^2 - \frac{1}{g^{tt}}\Bigg\}^{1/2},
\end{equation}
and its plot for $+L$ is in Fig. \ref{fig5}. The vertical dotted line is the event horizon for $a=0.50m$. The dash-dotted black line is the Schwarzschild case $k=a=0$. We verified that the curves (even at lower energies) never enter the event horizon. We can see how the spin parameter $a$ affects the maxima of the effective potential curve, which represents the radius for the unstable circular orbit. In such a peak, the deviation to the peak energy due to the cored profile is nearly identical to the Kerr case. With the same parameters, however, the cuspy profile increases further the peak energy where the unstable circular orbit occurs. The plot also shows that the energy at which a stable circular orbit occurs, which is represented by the minima of the curve, is highest in the cuspy profile. Furthermore, it happens at a lower value of $r/m$ than in the cored profile. As for the bound elliptical orbits, the cuspy profile admits a lower range for $r/m$ than in the cored profile.

\begin{figure}
    \centering
    \includegraphics[width=0.60\textwidth]{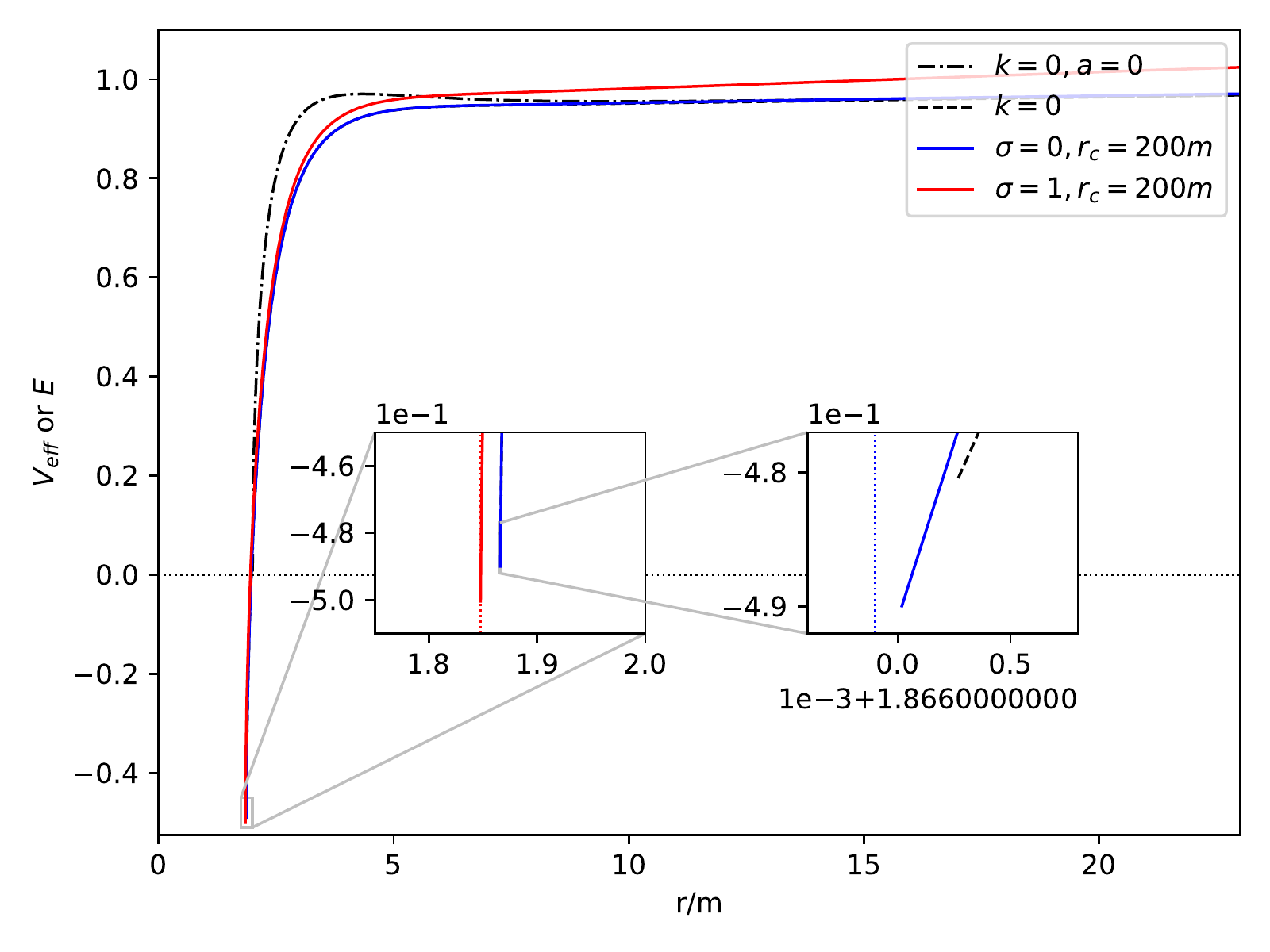}
    \caption{The effective potential when $a=0.50m$ for $-L$. The value of the angular momentum per unit mass is $L=-3.75m$.}
    \label{fig6}
\end{figure}

The energy extraction from a black hole is useful when the angular momentum per unit mass is negative. It is known that this is not allowed in the Schwarzschild case. In Fig. \ref{fig6}, we can see that any massive particle that counter-rotates the black hole will eventually spiral into the event horizon after reaching a potential wall. The energy of this potential wall is greater in the cuspy profile than in the cored profile.

Looking at the values of the particles with negative energy, it still occurs outside the event horizon (see first inset plot). By observing the two bottom inset plots, we can conclude that the effect of dark matter is to decrease the value of the radial distance from the black hole where the Penrose process occurs. The cored profile shows indistinguishable deviation from the Kerr case but offers slightly lower negative energies than the Kerr case. The cuspy profile shows more deviation and allows more particles with higher negative energies to be produced.

\section{Null geodesics} \label{sec6}
\begin{figure}
    \centering
    \includegraphics[width=0.60\textwidth]{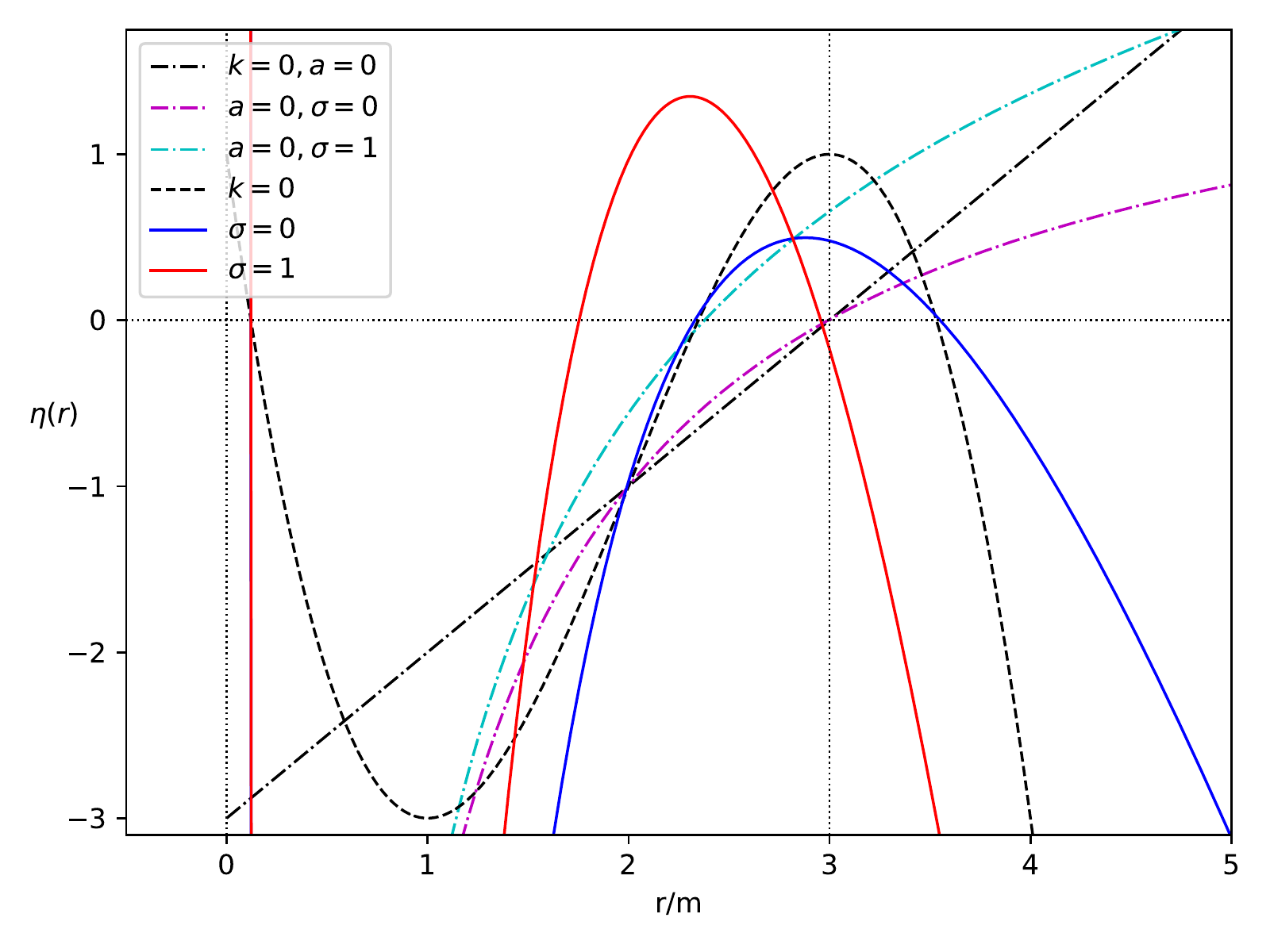}
    \caption{Plot of $\eta(r)=0$ showing the location of the photonspheres. Dash-dotted line is for the Schwarzschild case, the dashed line is for the Kerr case, and the solid line is for the Kerr case in Dehnen profile.}
    \label{fig7}
\end{figure}

\begin{figure}
    \centering
    \includegraphics[width=0.60\textwidth]{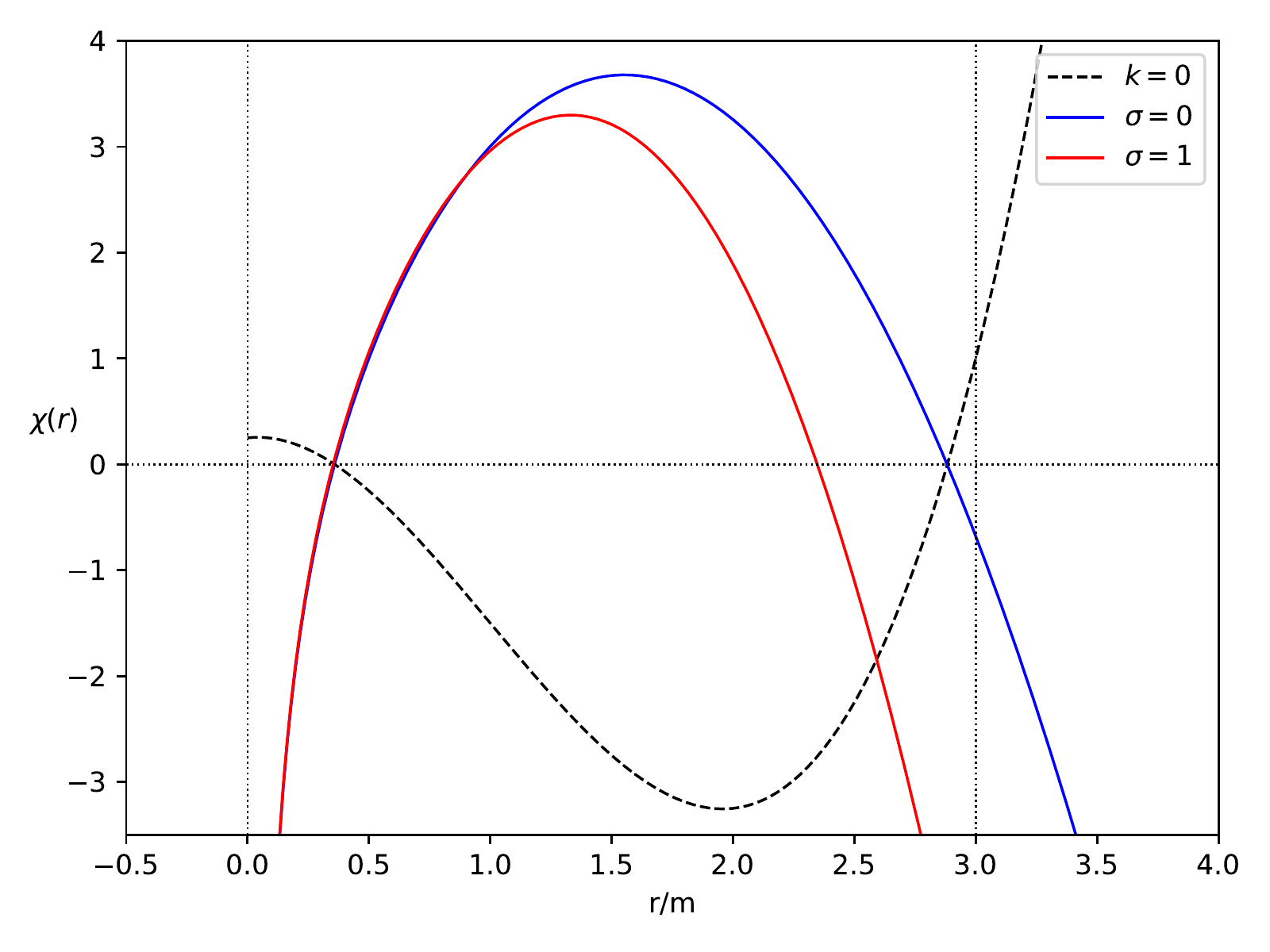}
    \caption{Plot of $\xi(r)=0$ showing the location of the photonspheres for zero angular momentum case when $a=0.50m$.}
    \label{fig8}
\end{figure}

Null geodesics are important in the study of the black hole shadow. Unlike a massive particle that can have a stable circular orbit, a photon always has an unstable circular orbit. Such an orbit determines the what would be the shape of the black hole shadow. The shadow contour, as perceived by a remote static observer, can be plotted using backward ray tracing. For photons, we set $\mu=0$, and it is useful to define the following impact parameters:
\begin{equation} \label{e45}
    \xi=\frac{L}{E}, \quad \eta=\frac{Q}{E^2}.
\end{equation}
Using this to $R(r)$ in Eq. \eqref{e38}, and after some algebra using the condition in Eq. \eqref{e39}, we find
\begin{equation} \label{e46}
\xi=\frac{\Delta'(r)(r^{2}+a^{2})-4\Delta(r)r}{a\Delta'(r)},
\end{equation}
\begin{equation} \label{e47}
\eta=\frac{-r^{4}\Delta'(r)^{2}+8r^{3}\Delta(r)\Delta'(r)+16r^{2}\Delta(r)(a^{2}-\Delta(r))}{a^{2}\Delta'(r)^{2}}.
\end{equation}
The location of the photon's unstable orbit can be found when we set $\eta(r)=0$, where analytic solutions are available for both the Schwarzschild and Kerr cases. However, since $\Delta(r)$ is complicated in this study, we analyze Eq. \eqref{e47} numerically, keeping in mind that we set $k=100m, r_\text{c} = 80m$ to illustrate the Dehnen profile's effect theoretically. In Fig. \ref{fig7}, the dash-dotted black line is the Schwarzschild case, and the photonsphere radius is at $r=3m$ as expected. We also included the dark matter effect in the Schwarzschild case, where the cuspy profile shows more deviation than the cored profile. For the Kerr case, represented by the black dashed line, we see two common values for the photonsphere radius. The closer one is for the prograde case $r = 2.35m$, and the farther one is for the retrograde case $r = 3.53m$. Respectively, for the extreme case, it is known that these radii are at $r=m$ and $r=4m$. When the black hole is surrounded by dark matter in the Dehnen profile, we see a differing behavior for the cored and cuspy profiles. While the cored profile's effect is tiny, its effect is to decrease the prograde orbit slightly while increasing the retrograde orbit of the photon. The cuspy profile's effect, however, drastically decreases the photon radii for both cases, given the same parameters herein. Lastly, the maxima in the plot represent the situation where the time derivative of $\phi$ is zero, which means that photons travel vertically as it crosses the equatorial plane. The radius where this happens decreases due to the Dehnen profile, where the cuspy profile shows the largest decrease.
For completeness, we also plotted the photon radius for zero angular momentum by setting $\xi(r)=0$. Along with the condition $\eta>0$, it is the only radius where photons can reach the poles at $\theta=0$ and $\theta=\pi$. Based on Fig. \ref{fig8}, the cored profile increases such radius, but such a deviation is very small relative to the Kerr case. The cuspy profile shows more deviation, and it tends to decrease the radius. The inner radius, which is inside the null boundaries, is also affected. In this region, the deviation for both profiles is very small. Nevertheless, the cored profile decreases such radius, while the cuspy profile increases its value. It is unknown what happens inside the null boundaries.
\begin{table} [!ht]
    \centering
    \begin{tabular}{ |c|c|c| }
    \hline
    Profile &  Prograde & Retrograde \\
    \hline
    no DM ($k=0$) & 2.34729635533386069770  & 3.53208888623795607040   \\
    cored ($\sigma=0$) & 2.34729635533386069770 & 3.53208888623795607040  \\
    cuspy ($\sigma=1$) & 2.34729635533386068830  & 3.53208888623795606100   \\
    \hline
    \end{tabular}
    \caption{Photonsphere radii due to a black hole in Leo I immersed in dark matter halo described by the Dehnen profile.}
    \label{Tab4}
\end{table}
\begin{table} [!ht]
    \centering
    \begin{tabular}{ |c|c|c| }
    \hline
    Profile &  Prograde & Retrograde \\
    \hline
    no DM ($k=0$) & 0.35858806834560123066  & 2.88321774192635239270   \\
    cored ($\sigma=0$) & 0.35858806834560123066 & 2.88321774192635239270  \\
    cuspy ($\sigma=1$) & 0.35858806834560123060  & 2.88321774192635238390   \\
    \hline
    \end{tabular}
    \caption{Photonsphere radii for zero angular momentum case due to a black hole in Leo I immersed in dark matter halo described by the Dehnen profile.}
    \label{Tab5}
\end{table}
To complete the treatment for null geodesics, we considered the application of the Dehnen profile for Leo I. Relative to Fig. \ref{fig7}, Table \ref{Tab4} shows that the deviation in both prograde and retrograde cases occurs around $10^{-17}$ order of magnitude.Relative to Fig. \ref{fig8}, Table \ref{Tab5} while the retrograde case occurs at $10^{-17}$ order of magnitude, the prograde case is at $10^{-19}$. It reveals that the Dehnen profile has an effect that is vanishingly small inside the null boundaries.

\section{Black hole shadow} \label{sec7}
\begin{figure}
    \centering
    \includegraphics[width=0.475\textwidth]{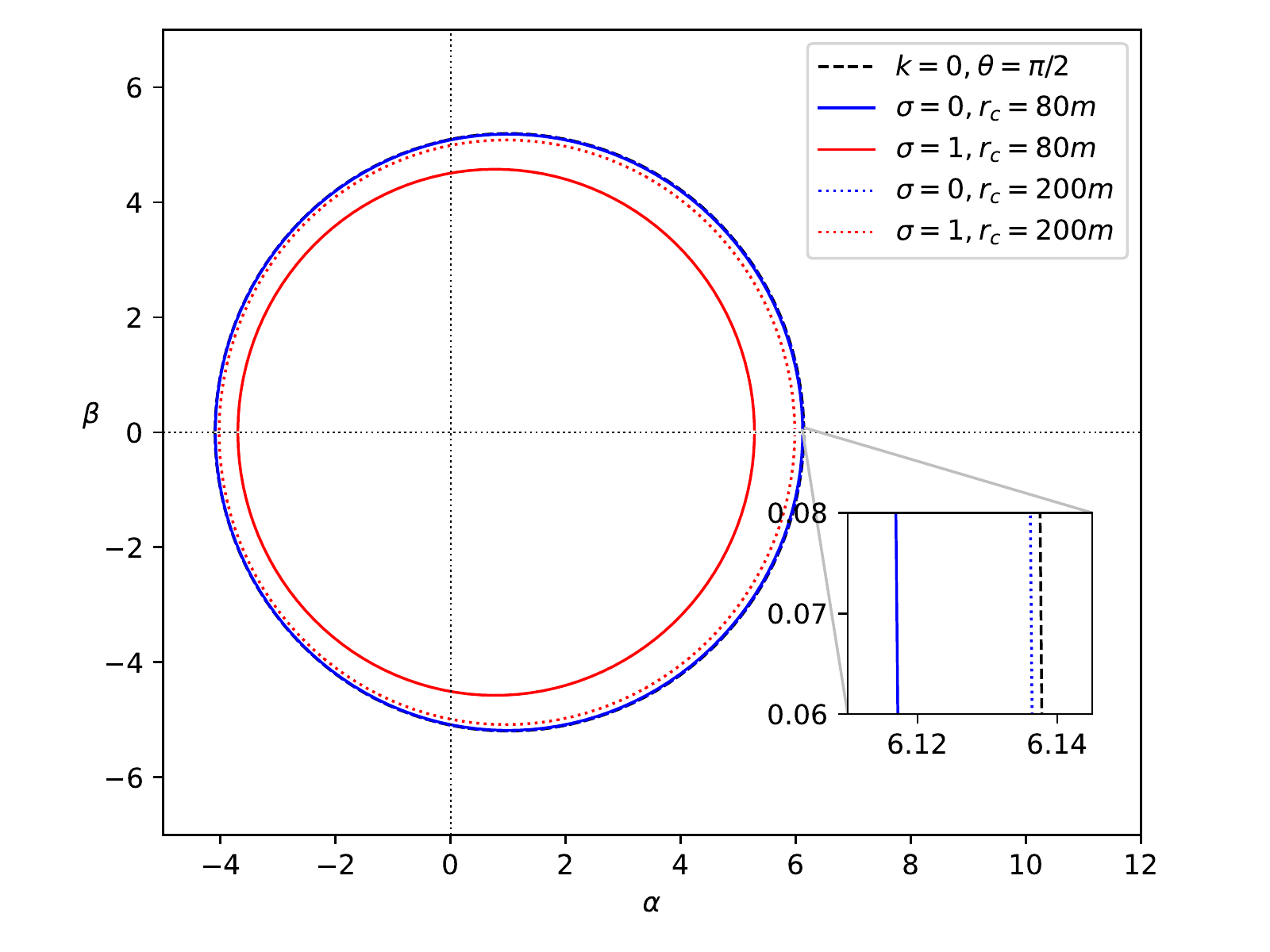}
    \includegraphics[width=0.475\textwidth]{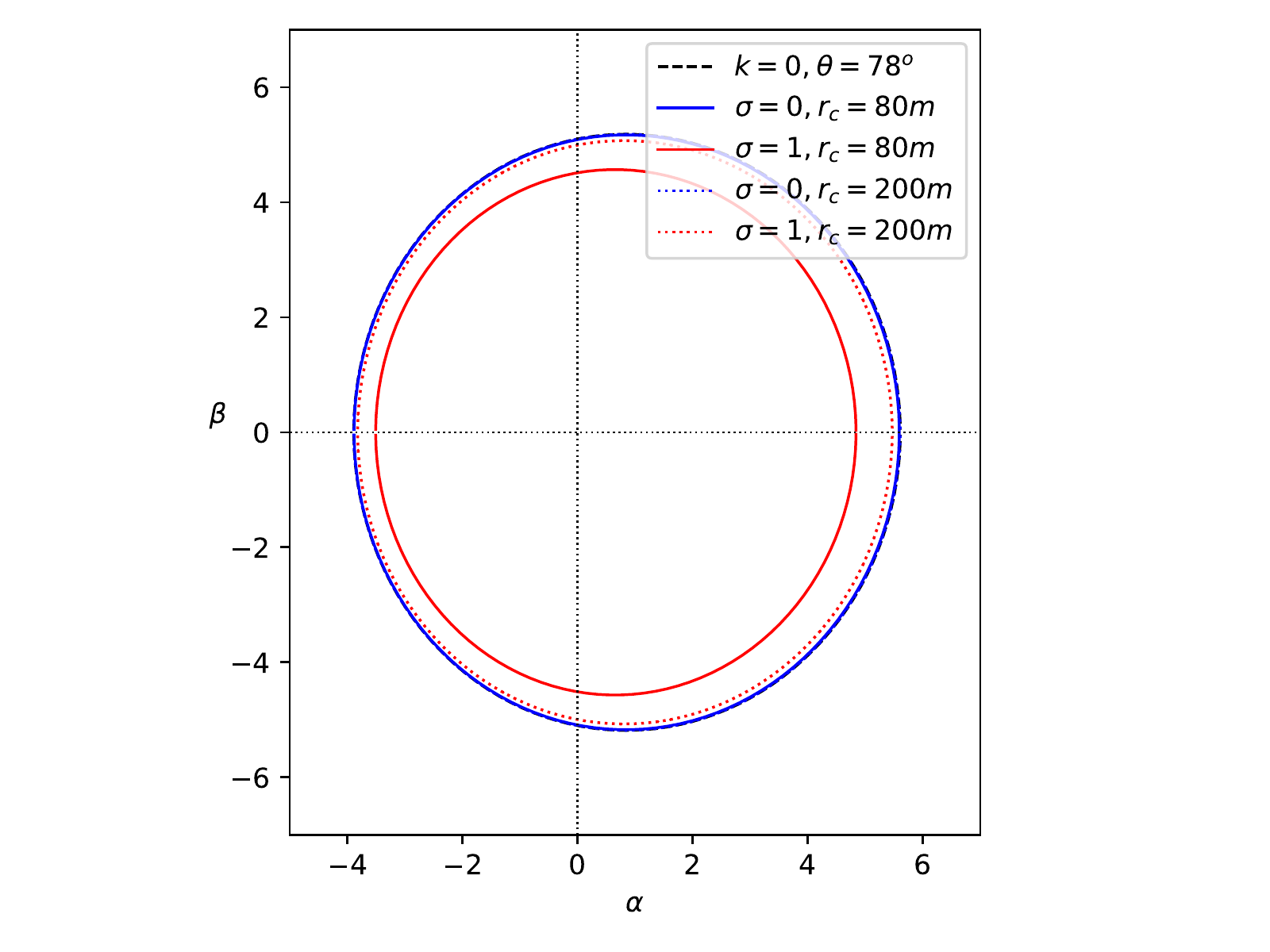}
    \caption{The black hole shadow with the Dehnen profile for $a=0.50m$ (left). The figure on the right shows how the shadow cast viewed $12^o$ declination.}
    \label{fig9}
\end{figure}

\begin{figure*}
    \centering
    \includegraphics[width=0.475\textwidth]{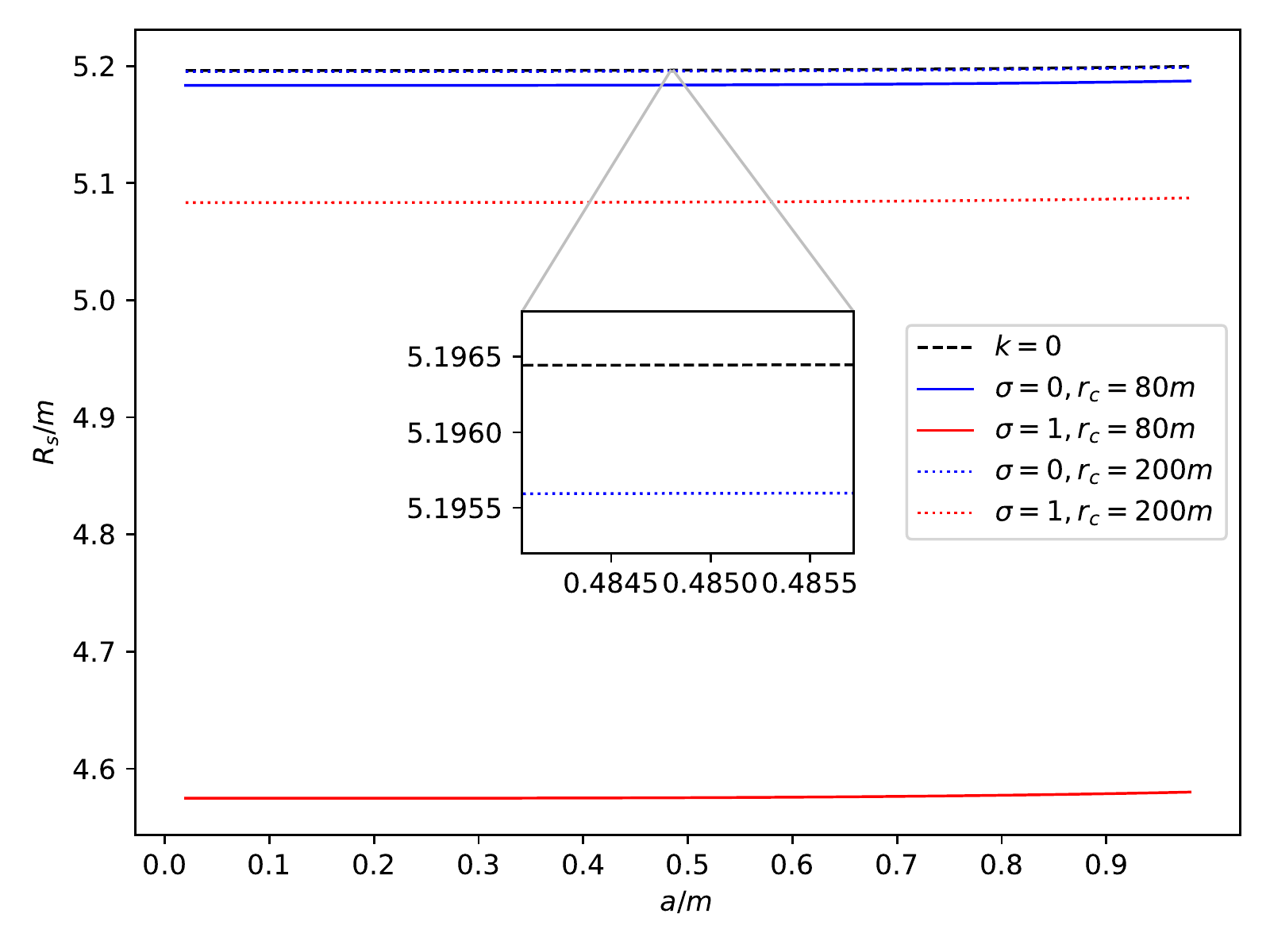}
    \includegraphics[width=0.475\textwidth]{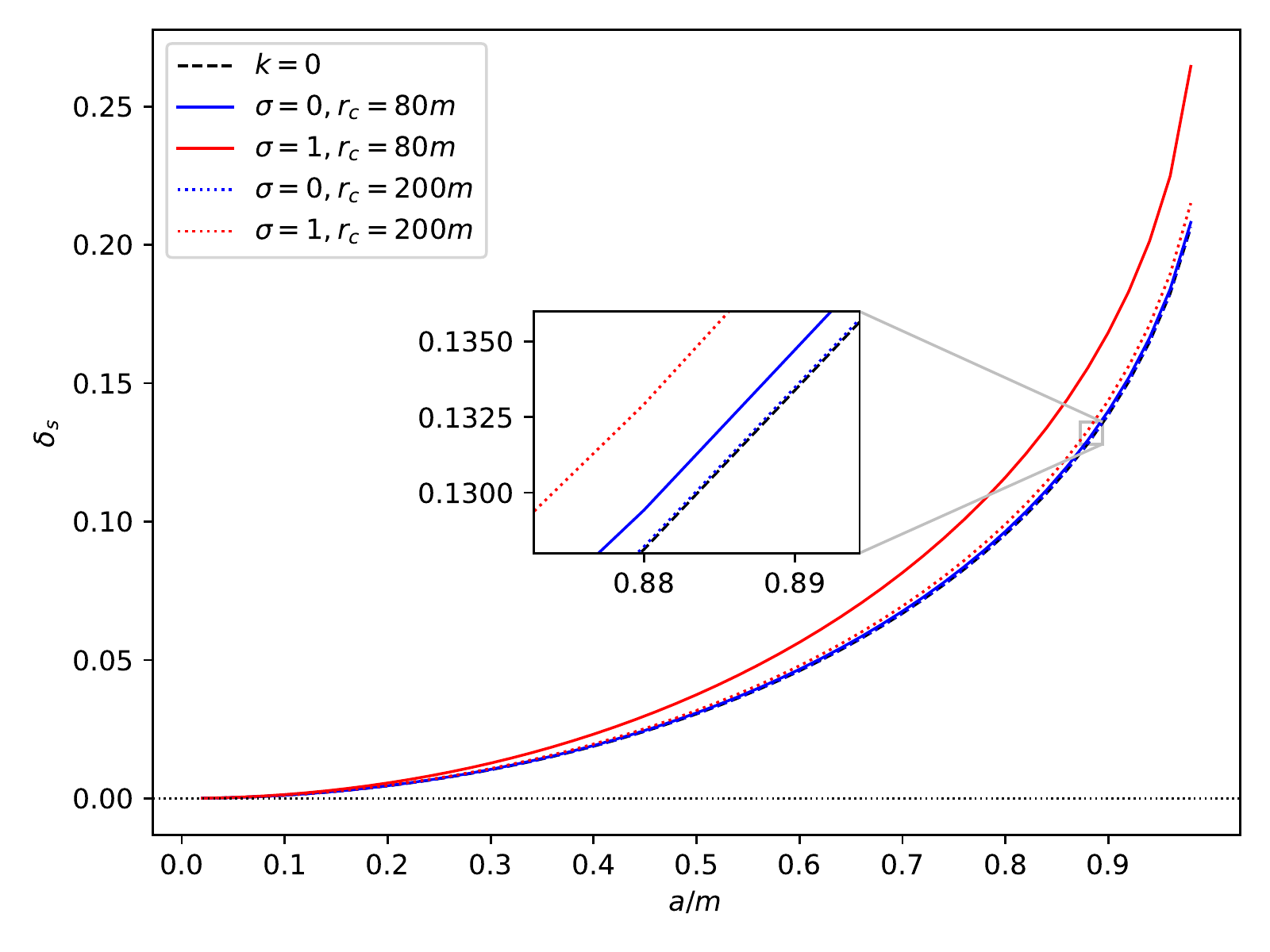}
    \caption{Plot of the shadow radius and distortion parameter.}
    \label{fig10n}
\end{figure*}

Any perturbation can cause the photons in the unstable orbit to escape or plunge into the black hole. For such reason, it is possible for an observer to form a shadow cast using the celestial coordinates are $(r_{\text{o}},\theta_{\text{o}})$. Such an observer is called the Zero Angular Momentum Observer (ZAMO), and the approximation $r_{\text{o}} \rightarrow \infty$ is taken, while $\theta_{\text{o}} = \pi/2$. The celestial coordinates are defined as \cite{Johannsen2013}
\begin{align} \label{e48}
    \alpha  &=-r_{\text{o}}\frac{\xi }{\zeta\sqrt{g_{\phi \phi }} \left( 1+\frac{g_{t\phi }}{g_{\phi \phi }}\xi\right) }, \nonumber\\
    \beta  &= r_{\text{o}}\frac{\pm \sqrt{\Theta (i)}}{\zeta\sqrt{g_{\theta \theta }} \left( 1+\frac{g_{t\phi}}{g_{\phi \phi }}\xi \right) },
\end{align}
and when $r_{\text{o}} \rightarrow \infty$, it is simplified as
\begin{align} \label{e49}
    \alpha&=-\xi \csc \theta_{\text{o}},   \nonumber \\
    \beta&=\pm \sqrt{\eta +a^{2}\cos ^{2}\theta_{\text{o}}-\xi ^{2}\cot^{2}\theta_{\text{o}}}.
\end{align}
Note how the above expression simplifies when $\theta_{\text{o}}=\pi/2$, and if $a=0$, the shadow cast of a Schwarzschild black hole is a circle. The plot of $\beta$ vs. $\alpha$ is shown in Fig. \ref{fig9} as $a=0.50m$ as well as the plot of the shadow radius as $a/m$ varies, which can be calculated using
\begin{equation} \label{e51n}
    R_\text{s}=\frac{\beta_{\text{t}}^2+(\alpha_{\text{t}}-\alpha_{\text{r}})^2}{2|\alpha_{\text{t}}-\alpha_{\text{r}}|}.
\end{equation}
For the figure associated with the variables in Eq. \eqref{e51n}, see Ref. \cite{Pantig:2020uhp}. For its derivation, see Ref. \cite{Dymnikova2019}. 
The black dashed line is the Kerr case. Note that the D-shaped shadow contour becomes more evident if $a=m$, but since we only deal with the non-extremal case, the contour is somewhat circular but distorted. In the inset plot, the cored profile gives little deviation from the Kerr case, and the cuspy profile shows more deviation, especially if the dark matter density is very high (solid line). Both profiles' overall effect is to decrease the size of the shadow. These statements are also clearly seen in the shadow radius plot on the right of Fig. \ref{fig10n}. As a final remark, we can now see the difference between the cored and cuspy profiles by comparing Figs. \ref{fig7} and \ref{fig9}. Note that the photonsphere of retrograde photons increases in the cored profile, but as these photons travel in the intervening space immersed in dark matter, the shadow size decreases. For the cuspy profile, both the photonsphere and shadow size decrease. Thus, based on this result, the behavior in the photonsphere, whether it increases or decreases, will not dictate what happens to the size of the shadow. The nature of dark matter described by the Dehnen profile causes the shadow size to decrease.

Another observable that can be derived from the shadow is the distortion parameter given by
\begin{equation}
    \delta_\text{s}=\frac{d_{\text{s}}}{R_{\text{s}}}=\frac{\tilde{\alpha}_{\text{l}}-\alpha_{\text{l}}}{R_{\text{s}}}.
\end{equation}
\begin{table} [!ht]
    \centering
    \begin{tabular}{ |c|c| }
    \hline
    Profile &  Shadow radius \\
    \hline
    no DM ($k=0$) & 5.19647974024604594750   \\
    cored ($\sigma=0$) & 5.19647974024604594750  \\
    cuspy ($\sigma=1$) & 5.19647974024604589700  \\
    \hline
    \end{tabular}
    \caption{Shadow radii due to a black hole in Leo I immersed in dark matter halo described by the Dehnen profile.}
    \label{Tab6}
\end{table}
For Leo I, we tabulated in Table \ref{Tab6} that shadow radius' numerical values. The cored profile shows an identical result to the Kerr case, but the cuspy profile shows deviation at $10^{-16}$ order of magnitude.

In the right of Fig. \ref{fig10n} we plot $\delta_\text{s}$ vs. $a/m$. For the Kerr case, we can only see an evident increase in the distortion parameter near $a=m$. The decrease in the shadow size caused by the dark matter effect is accompanied by higher values of distortion parameter for a given $a=m$, where the cuspy profile gives the highest distortion.

\begin{figure}
    \centering
    \includegraphics[width=0.60\textwidth]{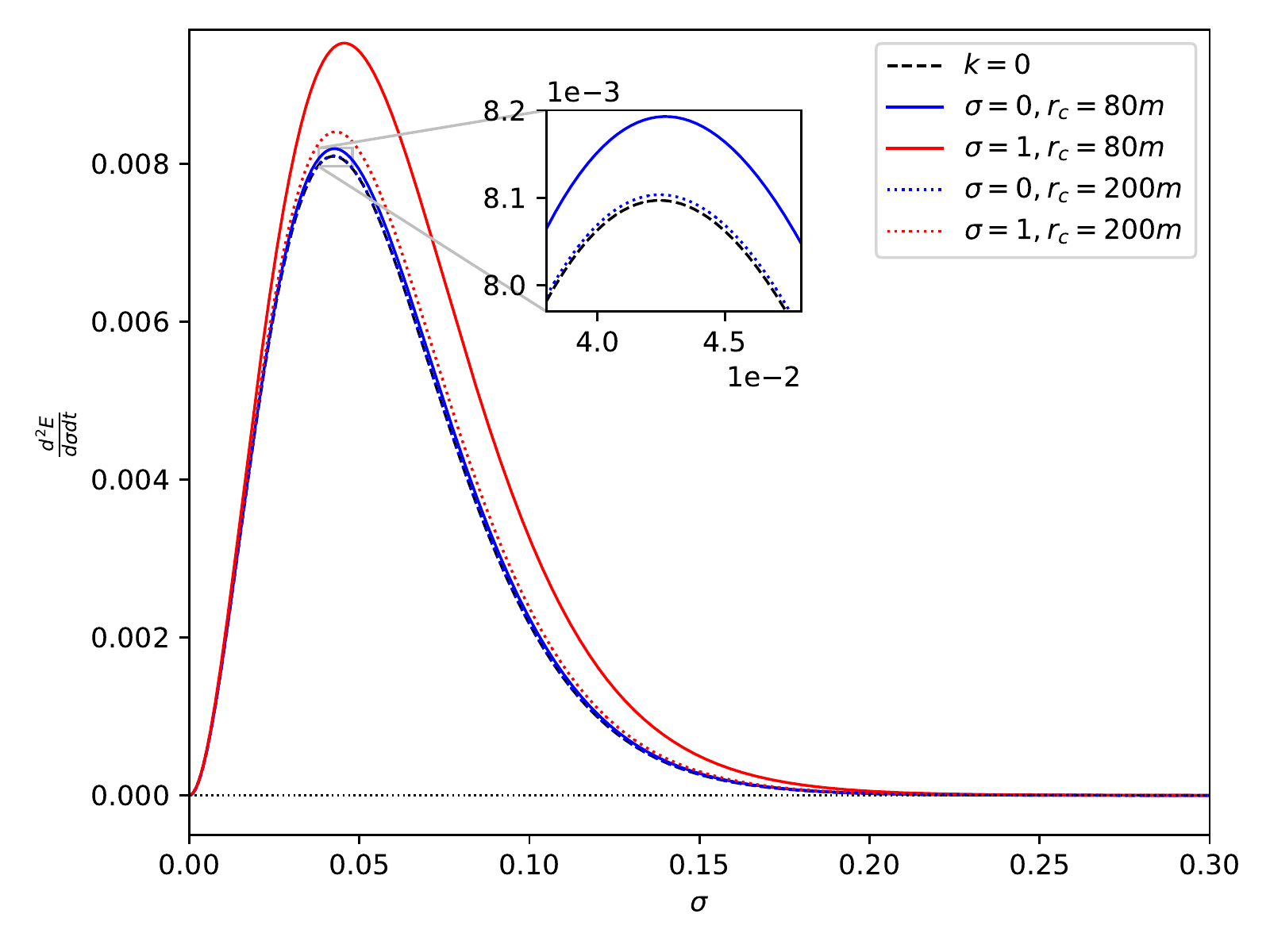}
    \caption{Energy emission rate for $a=0.50m$.}
    \label{fig12n}
\end{figure}

Finally, we also considered the black hole's energy emission rate defined by
\begin{equation} \label{e54n}
\frac{d^{2}E}{d\sigma dt}=2\pi^{2}\frac{\Pi_{ilm}}{e^{\sigma'/T}-1}\sigma'^{3},
\end{equation}
where the energy absorption cross-section $\Pi_{ilm} \sim \pi R_{\text{s}}^2$ for an observer at $r_\text{o} \rightarrow \infty$. Moreover, $T$ is the black hole temperature, which can be derived using the formula \cite{Jusufi2019}
\begin{equation} \label{e55n}
T=\frac{r_{\text{h}}}{4\pi(r_{\text{h}}^{2}+a^{2})^{2}}\left[2a^{2}(f(r_{\text{h}})-1)+r_{\text{h}}(r_{\text{h}}^{2}+a^{2})f'(r_{\text{h}})\right],
\end{equation}
where $r_{\text{h}}$ is the radius of the event horizon, and $f(r_{\text{h}})=g_{tt}$ in Eq. \eqref{e29} 
In Fig. \ref{fig12n}, we plot the energy emission rate vs the frequency $\sigma'$ for $a=0.50m$ only since we are interested on the dark matter effects.
The dark matter effect in the Dehnen profile tends to increase the energy emission rate of the black hole, showing the cuspy profile to have a greater deviation than the cored profile. It is also clear that for higher dark matter density, the higher the frequency of these emitted particles. Thus, a black hole in a cuspy profile has a shorter lifetime than a black hole surrounded by dark matter in cored profile.

\section{Weak deflection angle} \label{sec8}
In this section, we aim to calculate the weak deflection angle of the spinning black hole in the Dehnen dark matter halo using the combined methods in Refs. \cite{Ono:2019hkw} and \cite{Li:2020wvn}. The main reason is that the rotating metric in Eq. \eqref{e29} is a non-asymptotically flat spacetime. It is worth noting that the methods in Refs. \cite{Ono:2019hkw}, and \cite{Li:2020wvn} is based on the Gauss-Bonnet theorem, where modifications are made to accommodate non-asymptotically flat spacetimes. Let us introduce these methods briefly.

The Gauss-Bonnet theorem states that \cite{Carmo2016,Klingenberg2013}
\begin{equation} \label{e50}
    \iint_DKdS+\sum\limits_{a=1}^N \int_{\partial D_{a}} \kappa_{\text{g}} d\ell+ \sum\limits_{a=1}^N \theta_{a} = 2\pi\chi(D).
\end{equation}
Here, $D$ is the regular domain of a freely orientable two-dimensional curved surface $S$, which is described by the Gaussian curvature $K$. The infinitesimal area element is $dS$. The boundary of $D$ is given by $\partial D_{\text{a}}$ (a=$1,2,..,N$), and the geodesic curvature is integrated over the $d\ell$. Also, $\theta_\text{a}$ is the external angle, which $\chi(D)$ is the Euler characteristic, which in our case, is equal to $1$. Consider the positions of the lensing object, which is the spinning black hole, the source of light S, and the receiver R. In the axisymmetric case, the key equation
\begin{equation} \label{e51}
    \hat{\alpha}=\Psi_{\text{R}}-\Psi_{\text{S}}+\phi_{\text{RS}}
\end{equation}
becomes \cite{Ono:2017pie},
\begin{equation} \label{e52}
    \hat{\alpha} = -\iint_{_{\text{R}}^{\infty }\square _{\text{S}}^{\infty}}KdS-\int_{\text{S}}^{\text{R}} \kappa_{\text{g}}d\ell,
\end{equation}
which is valid for asymptotically flat spacetimes. However, for non-asymptotically flat spacetime, the authors in Ref. \cite{Ono:2019hkw} have shown that
\begin{equation} \label{e53}
    \hat{\alpha}=\Psi_{\text{R}}-\Psi_{\text{S}}+\int_{u_{R}}^{u_{o}}\frac{1}{\sqrt{F(u)}}du+\int_{u_{S}}^{u_{o}}\frac{1}{\sqrt{F(u)}}du,
\end{equation}
can be used even for axisymmetric cases. In the above equation, we can see how $\phi_{\text{RS}}$ should be calculated. In particular, one must use the orbit equation given as
\begin{equation} \label{e54}
    F(u)=u^4\frac{(A(u)D(u)+H(u)^2)(D(u)-2H(u)b-A(u)b^2)}{B(u)(H(u)+A(u)b)^2}
\end{equation}
and integrate it. Here, the orbit equation is expressed in terms of $u=1/r$. The integration is such a daunting task, especially with $A(r)$ in Eq. \eqref{e13}. However, in Ref. \cite{Li:2020wvn} it was shown how $\phi_{\text{RS}}$ should be alternatively calculated using the finite version of the Gauss-Bonnet theorem. Using the circular orbit of either the null or time-like particle, it was shown that
\begin{equation} \label{e55}
    \hat{\alpha} = \iint_{_{r_{\text{co}}}^{R }\square _{r_{\text{co}}}^{S}}KdS + \phi_{\text{RS}},
\end{equation}
where $\phi_{\text{RS}}$ is found through solving $F(u)$ by iteration. The aim is to solve for the closest approach $u_{\text{o}}$, then solve for the angle $\phi$. Although Eq. \eqref{e55} is not on its axisymmetric generalization, what matters here is how $\phi_{\text{RS}}$ is calculated. It is still tedious, but the method avoids integration.

\begin{figure}
    \centering
    \includegraphics[width=0.6\textwidth]{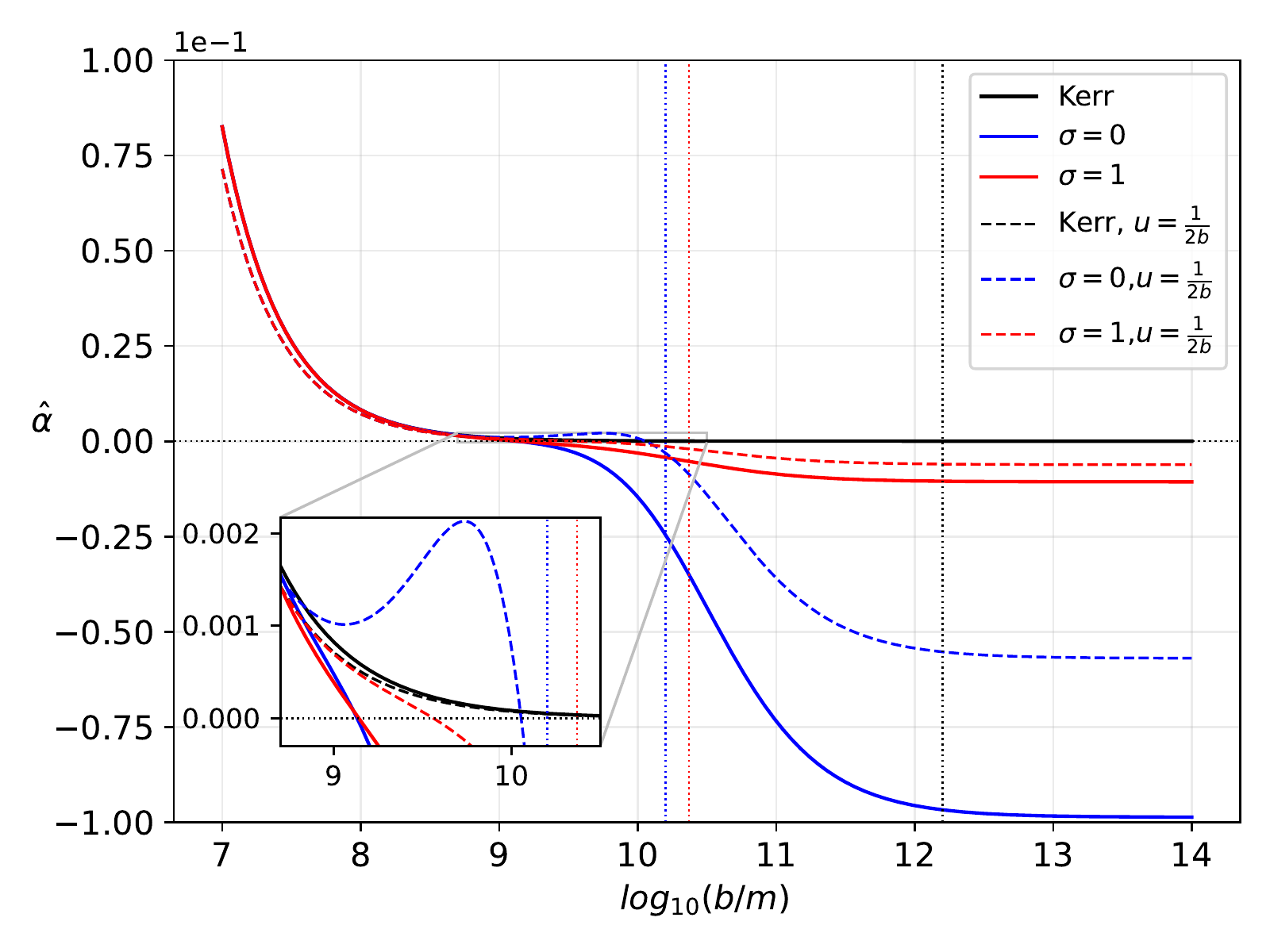}
    \caption{The weak deflection angle by a black hole in the Dehnen profile where the black hole spin parameter is $a=0.50m$. Here, $\hat{\alpha}$ is in $\mu$as. Leo I distance $250$ kpc \cite{Karachentsev:2004dx,2006Ap.....49....3K}. Due to the negligible effect of the Dehnen profile, the result for $a = -0.50m$ is nearly the same. We also indicate the core radius represented by the vertical dotted lines color coded to the profile. Note that the black vertical line corresponds to the distance of Earth from Leo I.}
    \label{fig10}
\end{figure}

Now using the rotating metric in Eq. \eqref{e29}, we find the orbit equation as
\begin{align} \label{e56}
    F(u) &= \frac{1}{b^{2}} - u^{2} + 2mu^{3} -\frac{4amu}{b^{3}}+\frac{2ku^{2}(r_{\text{c}} u+1)^{\sigma-2}}{r_{\text{c}}(\sigma-2)}-\frac{4ak(r_{\text{c}} u+1)^{\sigma-2}}{b^{3}r_{\text{c}}(\sigma-2)} \nonumber\\
    &-\frac{16akmu(r_{\text{c}} u+1)^{\sigma-2}}{b^{3}r_{\text{c}}(\sigma-2)} + \mathcal{O}(m^2,k^2,a^2m^2,a^2k^2,a^2k^2m^2).
\end{align}
We see the leading orders for the couplings terms $am$, $ak$, and $amk$. Note that the dark matter contribution stills gives some restriction for $\sigma$ and Eq. \eqref{e56} is only valid for the interval $[0,2)$. Furthermore, those terms with the spin parameter $a$ tend to decrease the value for $F(u)$. Noting the $F(u) = (du/d\phi)^2$, we differentiate Eq. \eqref{e56} again and obtain
\begin{equation} \label{e57n}
    \frac{1}{2}kr_{\text{c}}(\sigma-2)u^{2}(r_{\text{c}} u+1)^{\sigma-3}+ku(r_{\text{c}} u+1)^{\sigma-2}+\frac{1}{2}r_{\text{c}}(\sigma-2)\left(3mu^{2}-u-\frac{d^{2}u}{d\phi^{2}}\right)=0.
\end{equation}
Indeed, the above equation is a second-order non-linear differential equation. However, we can use perturbative methods \cite{Casana2018} to find an approximate solution for $u$. Here, we only consider the linear terms on Eq. \eqref{e57n}. That is, the differential equation
\begin{equation}
    -\frac{1}{2}r_{\text{c}}(\sigma-2)\left(u+\frac{d^{2}u}{d\phi^{2}}\right)=0
\end{equation}
must be solved, which gives
\begin{equation}
    u=X\sin(\phi)+Y\cos(\phi)
\end{equation}
where $X$ and $Y$ and constants that need to be determined through boundary conditions. Since $\frac{du}{d\phi}\big|_{\phi=\frac{\pi}{2}}=0$,
we find that the first term only matters, and $X = 1/b$. Therefore,
\begin{equation}
    u_{\text{o}}=\frac{1}{b}\sin(\phi)
\end{equation}
which is the initial expression to be used in the iteration method. Noting again Eq. \eqref{e56}, we already know at least four terms in the solution for the closest approach $u_{\text{o}}$. Proceeding with the process \cite{Ono:2019hkw,Li:2020wvn}, we finally obtained
\begin{align} \label{e57}
    u_{\text{o}}&=\frac{1}{b}\sin(\phi)+\frac{m(1+\cos^2(\phi))}{b^{2}}-\frac{2am}{b^{3}}+\frac{k}{br_{\text{c}}(\sigma-2)}\left(1+\frac{r_{\text{c}}}{b}\right)^{\sigma-2}-\frac{2ak}{b^{2}r_{\text{c}}(\sigma-2)}\left(1+\frac{r_{\text{c}}}{b}\right)^{\sigma-2} \nonumber\\
    &-\frac{4akm(4b+r_{\text{c}}\sigma+2r_{\text{c}})}{b^{3}r_{\text{c}}(\sigma-2)(b+r_{\text{c}})}\left(1+\frac{r_{\text{c}}}{b}\right)^{\sigma-2} + \mathcal{O}(m^2,k^2,a^2m^2,a^2k^2,a^2k^2m^2),
\end{align}
which can be used to solve for $\phi_{\text{RS}}$ \cite{Li:2020wvn}. We find $\phi$ as
\begin{align} \label{e58}
    \phi&=\arcsin(bu)+\frac{m}{b}\frac{b^2u^2-2}{\sqrt{-b^2u^2+1}}+\frac{2ma}{b^{2}}\frac{1}{\sqrt{1-b^{2}u^{2}}}-\frac{k}{r_{\text{c}}(\sigma-2)}\left(1+\frac{r_{\text{c}}}{b}\right)^{\sigma-2}\frac{1}{\sqrt{1-b^{2}u^{2}}} \nonumber\\
    &-\frac{b^{2}km}{r_{\text{c}}(\sigma-2)}\left(1+\frac{r_{\text{c}}}{b}\right)^{\sigma-2}\frac{u^{3}}{\sqrt{1-b^{2}u^{2}}\left(b^{2}u^{2}-1\right)}+\frac{2ak}{br_{\text{c}}(\sigma-2)}\left(1+\frac{r_{\text{c}}}{b}\right)^{\sigma-2}\frac{1}{\sqrt{1-b^{2}u^{2}}} \nonumber\\
    &+\frac{2akm}{b^{2}r_{\text{c}}(\sigma-2)(b+r_{\text{c}})}\left(1+\frac{r_{\text{c}}}{b}\right)^{\sigma-2}\frac{\gamma}{\sqrt{1-b^{2}u^{2}}\left(b^{2}u^{2}-1\right)} + \mathcal{O}(m^2,k^2,a^2m^2,a^2k^2,a^2k^2m^2),
\end{align}
where
\begin{equation}
    \gamma=b^{4}u^{3}+b^{3}\left(r_{\text{c}} u^{3}+8u^{2}\right)+b^{2}\left[2r_{\text{c}}(\sigma+2)u^{2}+u\right]+b(r_{\text{c}} u-8)-2r_{\text{c}}(\sigma+2).
\end{equation}
Note that in the above equation, $\phi_{\text{R}}=\pi-\phi|_{u\rightarrow u_{\text{R}}}$ and $\phi_{\text{S}}=\phi|_{u\rightarrow u_{\text{S}}}$. Thus, the expression for the longitudinal angle is found by using \cite{Li:2020wvn}
\begin{equation} \label{e59}
    \phi_{\text{RS}}=\phi_{\text{R}}-\phi_{\text{S}}.
\end{equation}
We do not show the expression here since it is rather lengthy. Next, we turn our attention to the positional angles $\Psi_\text{R}$ and $\Psi_\text{S}$. From the source to the receiver, the photons travel in the three-dimensional manifold where its unit tangent vector depends on the orbit equation \cite{Ono:2017pie}, ie.
\begin{equation}
	e^i=\frac{d\phi}{dt}\left(\frac{dr}{d\phi},0,1\right)
\end{equation}
where for the axisymmetric case,
\begin{equation}
	\frac{d\phi}{dt} = \frac{A(r)[H(r)+A(r)b]}{A(r)D(r)+H(r)^2}.
\end{equation}
As for the unit radial vector, taken positively in the outgoing direction, we have
\begin{equation}
	R^i=\left(\frac{1}{\sqrt{\gamma_{rr}}},0,0\right).
\end{equation}
With the definition $\cos \Psi = \gamma_{ij}e^iR^j$ and basic trigonometry, one can derive
\begin{equation} \label{e65n}
	\sin \Psi = \frac{H(r)+A(r)b}{\sqrt{A(r)D(r)+H(r)^2}}.
\end{equation}
With the use of Eq. \eqref{e65n},	we find
\begin{align} \label{e60}
    \Psi&=\arcsin(bu)-\frac{bmu^{2}}{\sqrt{1-b^{2}u^{2}}}-\frac{bk}{r_{\text{c}}(\sigma-2)}\frac{u(r_{\text{c}} u+1)^{\sigma-2}}{\sqrt{1-b^{2}u^{2}}}+\frac{bkm}{r_{\text{c}}(\sigma-2)}\frac{u^{2}\left(2b^{2}u^{2}-1\right)(r_{\text{c}} u+1)^{\sigma-2}}{\left(1-b^{2}u^{2}\right)^{3/2}} \nonumber\\
    &-\frac{2ak}{r_{\text{c}}(\sigma-2)}\frac{u(r_{\text{c}} u+1)^{\sigma-2}}{\sqrt{1-b^{2}u^{2}}}-\frac{4akm}{r_{\text{c}}(\sigma-2)}\frac{u^{2}\left(2b^{2}u^{2}-1\right)(r_{\text{c}} u+1)^{\sigma-2}}{\left(1-b^{2}u^{2}\right)^{3/2}} + \mathcal{O}(m^2,k^2,a^2m^2,a^2k^2,a^2k^2m^2).
\end{align}
Here, if one considers the finite distance of the source S and the receiver R, then \cite{Ono:2019hkw}
\begin{equation} \label{e61}
    \Psi_{\text{R}}-\Psi_{\text{S}} = \Psi|_{u\rightarrow u_{\text{R}}} - \pi + \Psi|_{u\rightarrow u_{\text{S}}}.
\end{equation}
Then with Eqs. \eqref{e59} and \eqref{e61}, the weak deflection angle can be computed as \cite{Ono:2019hkw}
\begin{equation} \label{e62}
    \hat{\alpha}=\Psi_{\text{R}}-\Psi_{\text{S}}+\phi_{\text{RS}}.
\end{equation}
The general equation for the weak deflection angle with finite distance is then
\begin{align} \label{e69n}
    \hat{\alpha}&=\frac{2m}{b}\left[\sqrt{1-b^{2}u_{\text{R}}^{2}}+\sqrt{1-b^{2}u_{\text{S}}^{2}}\right]\mp\frac{2am}{b^{2}}\left[\sqrt{1-b^{2}u_{\text{R}}^{2}}+\sqrt{1-b^{2}u_{\text{S}}^{2}}\right] \nonumber \\
    &+\frac{k}{r_{\text{c}}(\sigma-2)}\left\{ \frac{\left[\left(1+\frac{r_{\text{c}}}{b}\right)^{\sigma-2}-bu_{\text{R}}(r_{\text{c}} u_{\text{R}}+1)^{\sigma-2}\right]}{\sqrt{1-b^{2}u_{\text{R}}^{2}}}+\frac{\left[\left(1+\frac{r_{\text{c}}}{b}\right)^{\sigma-2}-bu_{\text{S}}(r_{\text{c}} u_{\text{S}}+1)^{\sigma-2}\right]}{\sqrt{1-b^{2}u_{\text{S}}^{2}}}\right\} \nonumber \\
    &-\frac{bkm}{r_{\text{c}}(\sigma-2)}\left\{ \frac{u_{\text{R}}^{2}\left[\left(1+\frac{r_{\text{c}}}{b}\right)^{\sigma-2}bu_{\text{R}}-(r_{\text{c}} u_{\text{R}}+1)^{\sigma-2}(2b^{2}u_{\text{R}}^{2}-1)\right]}{\left(1-b^{2}u_{\text{R}}^{2}\right)^{3/2}}+\frac{u_{\text{S}}^{2}\left[\left(1+\frac{r_{\text{c}}}{b}\right)^{\sigma-2}bu_{\text{S}}-(r_{\text{c}} u_{\text{S}}+1)^{\sigma-2}(2b^{2}u_{\text{S}}^{2}-1)\right]}{\left(1-b^{2}u_{\text{S}}^{2}\right)^{3/2}}\right\} \nonumber \\
    &\mp\frac{2ak}{br_{\text{c}}(\sigma-2)}\left\{ \frac{\left[\left(1+\frac{r_{\text{c}}}{b}\right)^{\sigma-2}+bu_{\text{R}}(r_{\text{c}} u_{\text{R}}+1)^{\sigma-2}\right]}{\sqrt{1-b^{2}u_{\text{R}}^{2}}}+\frac{\left[\left(1+\frac{r_{\text{c}}}{b}\right)^{\sigma-2}+bu_{\text{S}}(r_{\text{c}} u_{\text{S}}+1)^{\sigma-2}\right]}{\sqrt{1-b^{2}u_{\text{S}}^{2}}}\right\} \nonumber \\
    &\pm\frac{2akm}{b^{2}r_{\text{c}}(\sigma-2)(b+r_{\text{c}})}\Bigg\{ \frac{\left[\left(1+\frac{r_{\text{c}}}{b}\right)^{\sigma-2}\gamma_{\text{R}}-2b^{2}u_{\text{R}}^{2}(b+r_{\text{c}})\left(2b^{2}u_{\text{R}}^{2}-1\right)(r_{\text{c}} u_{\text{R}}+1)^{\sigma-2}\right]}{\left(1-b^{2}u_{\text{R}}^{2}\right)^{3/2}}+ \nonumber \\
    &\frac{\left[\left(1+\frac{r_{\text{c}}}{b}\right)^{\sigma-2}\gamma_{\text{S}}-2b^{2}u_{\text{S}}^{2}(b+r_{\text{c}})\left(2b^{2}u_{\text{S}}^{2}-1\right)(r_{\text{c}} u_{\text{S}}+1)^{\sigma-2}\right]}{\left(1-b^{2}u_{\text{S}}^{2}\right)^{3/2}}\Bigg\} + \mathcal{O}(m^2,k^2,a^2m^2,a^2k^2,a^2k^2m^2).
\end{align}
Here, the upper sign is used for the prograde orbit of the photon, while the lower sign is for the retrograde orbit. Assuming that $u_{\text{R}}=u_{\text{S}}$, which both approaches zero, and that the impact parameter $b$ is astronomically large, we obtain the weak deflection angle approximation as
\begin{equation} \label{e63}
    \hat{\alpha}=\frac{4m}{b}\mp\frac{4am}{b^{2}}+\frac{\left(1+\frac{r_{\text{c}}}{b}\right)^{\sigma-2}}{r_{\text{c}}(\sigma-2)}\left(2k\mp\frac{4ak}{b}\right) \mp\frac{\left(1+\frac{r_{\text{c}}}{b}\right)^{\sigma}}{r_{\text{c}}(\sigma-2)}\frac{8akm(4b+r_{\text{c}}\sigma+2r_{\text{c}})}{b^2(b+r_{\text{c}})}.
\end{equation}
We note that without the dark matter, we obtain the correct expression for the weak deflection angle by a Kerr black hole. Furthermore, if $a=0$, the Schwarzschild case is also obtained. In Fig. \ref{fig10}, we plot the weak deflection angle with finite distance Eq. \eqref{e69n} where $u=1/2b$ (dashed lines), and compare it with the weak deflection angle when $u\rightarrow 0$ described by Eq. \eqref{e63} (solid lines). The blue and red vertical dotted line represents the core radius $r_\text{c}$ of the dark matter profiles. The solid black line represents the weak deflection angle for the Kerr case without the influence of the dark matter environment, which is nearly zero when the impact parameter is very high. Its weak deflection angle, although of low value, becomes evident as $b/m$ decreases further (see inset plot). Generally, the Dehnen profile admits a negative deflection angle, which is not new since other black hole models also produce a negative deflection angle \cite{Nakashi2019,Pantig:2022toh}. The only consequence of $-\hat{\alpha}$ is its unobservability. Nonetheless, the weak deflection angles are positive within the core radius $r_\text{c}$. It is clear from the plot that the distance of the observer from the black hole (not only the impact parameter) affects the behavior of the weak deflection angle. Interestingly, the cored profile gives a differentiated behavior at finite distance near the dark matter's core radius, while the rest follow the trend of the Kerr case. For lower values of $b/m$, we can see that the weak deflection angle slightly decreased when finite distance is considered.
\begin{table} [!ht]
    \centering
    \begin{tabular}{ |c|c|c| }
    \hline
    Profile &  WDA in $\mu$as (Eq. \eqref{e69n}) & WDA in $\mu$as (Eq. \eqref{e63}) \\
    \hline
    no DM ($k=0$) & 0.02259519866476061761 & 0.02609068806298523043     \\
    cored ($\sigma=0$) & 0.02609029831699791768 & 0.02609029830687080475  \\
    cuspy ($\sigma=1$) & 0.02607630035135645743 & 0.02607630035026163438     \\
    \hline
    \end{tabular}
    \caption{Photonsphere radii for zero angular momentum case due to a black hole in Leo I immersed in dark matter halo described by the Dehnen profile. Here, $a=0.50m$ and $\log_{10}(b/m)=7.5$.}
    \label{Tab7}
\end{table}
\begin{table} [!ht]
    \centering
    \begin{tabular}{ |c|c|c| }
    \hline
    Profile &  WDA in $\mu$as (Eq. \eqref{e69n}) & WDA in $\mu$as (Eq. \eqref{e63}) \\
    \hline
    no DM ($k=0$) & 0.02259519866476061775 & 0.02609068806298523060     \\
    cored ($\sigma=0$) & 0.02609029914203329278 & 0.02609029913190618049  \\
    cuspy ($\sigma=1$) & 0.02607630117550652440 & 0.02607630117441170142     \\
    \hline
    \end{tabular}
    \caption{Photonsphere radii for zero angular momentum case due to a black hole in Leo I immersed in dark matter halo described by the Dehnen profile. Here, $a=-0.50m$ and $\log_{10}(b/m)=7.5$.}
    \label{Tab8}
\end{table}
We also include numerical values of $\hat{\alpha}$ for Leo I, where we both considered photons co- and counter-rotation cases. In both Tables \ref{Tab7} and \ref{Tab8}, the finite distance case is sensitive within $10^{-3}$ order of magnitude, and the far approximation case is at only $10^{-5}$ order of magnitude. Remarkably, while the Dehnen effect is difficult to detect using the deviations in the shadow cast, the weak deflection angle allows the deviation at lower orders of magnitude.

\section{Conclusion} \label{sec9}
With a few recent discoveries of black holes in dwarf galaxies, it motivated us to construct a black hole solution surrounded by dark matter whose profile fits this type of galaxy. In particular, we associated the Dehnen profile with a Schwarzschild black hole since such a profile is useful for ultra-faint dwarf galaxies. The equation governing the Dehnen profile can accommodate both cored $\sigma=0$ and cuspy $\sigma=1$ configurations. Acting as a seed metric, we utilized the modified Newman-Janis prescription to obtain the rotating solution to make a more realistic model. Then, we analyzed how the Dehnen profile affects the horizons, time-like and null orbits, the shadow and its observables, and the weak deflection angle with theoretical values of $k$ and $r_\text{c}$, and its application to the black hole in Leo I. Our results revealed that the cuspy profile shows more deviation to the usual black hole properties mentioned than the cored profile. The cuspy profile then leaves more room for detection since the cored profile is almost indistinguishable from the Kerr case. The study is different in that it uses a black hole instead of stellar distribution to determine how the cored and cuspy configuration of the Dehnen profile affects these properties.

Not usually done in the literature is the analysis of how the dark matter profile affects the orbits of a massive particle around a black hole. Compared to Dehnen's effect on the null boundaries, massive particles are more sensitive to its effect as shown by the deviations in ISCO, unstable and stable orbits, and the bound elliptical orbits. Deviation for the null orbits needs more dark matter concentration. In this study, we have shown how the shadow cast behaves under the Dehnen profile's effect. Its application to the black hole in Leo I reveals that such deviation in the shadow radius is vanishingly small. The deviation is even smaller than those reported in Refs. \cite{Xu2018,Jusufi2019,Jusufi2020,Jusufi2021,Nampalliwar2021,Konoplya2019,Pantig:2020odu,Pantig:2021zqe,Hou_2018a,Hou_2018b,Xu2021a,Xu2021b,Pantig:2020uhp}. We conclude that it is inherent to the Dehnen profile described by Eq. \eqref{e1}, and smaller black holes in dwarf galaxies may yield larger deviations. Detecting the deviation caused by the Dehnen profile on the shadow with current space technology like the EHT ($10\mu$as), ESA GAIA mission ($20\mu$as-$7\mu$as) \cite{Liu_2017}, and the futuristic VLBI RadioAstron that can achieve at around $1-10\mu$as \cite{Kardashev2013}, are not enough even if the dark matter is theoretically concentrated near the black hole.

The study has shown how difficult it is to determine whether the dark matter distribution in dwarf galaxies is cored or cuspy using the deviation in the shadow size. It is not the case if we determine this using the weak deflection angle phenomenon since we have shown that for Leo I, the deviations occur within $10^{-3}$ order of magnitude for the finite distance case. Consideration our location from Leo I, this sensitivity is reduced to only $10^{-5}$ (see Tables \ref{Tab7} and \ref{Tab8})

In light of this study, it is interesting to explore further the effect of the Dehnen profile on the weak and strong deflection angle of massive particles. Since time-like orbits are also affected, exploring their effect on a massive particle with a spin is also interesting. Another direction is to explore the Dehnen profile for small black holes within a dwarf galaxy. Finally, future studies with a spinning dark matter halo and a spinning black hole are also interesting.


\bibliography{references}
\bibliographystyle{apsrev}
\end{document}